\documentclass[conf]{new-aiaa}
\usepackage[utf8]{inputenc}

\usepackage{graphicx}
\usepackage{amsmath}
\usepackage[version=4]{mhchem}
\usepackage{siunitx}
\usepackage{longtable,tabularx}
\usepackage{color}
\usepackage{tikz}

\title{High-order curvilinear hybrid mesh generation \\ for CFD simulations}

\author{Julian Marcon \footnote{PhD Candidate, Department of Aeronautics. AIAA Member.},
             Michael Turner \footnote{PhD, Department of Aeronautics}, and 
            Joaquim Peiro\footnote{Reader, Department of Aeronautics, South Kensington Campus}}
\affil{Imperial College London, South Kensington Campus, London SW7 2AZ, United Kingdom}
\author{David Moxey\footnote{Lecturer, College of Engineering, Mathematics and Physical Sciences}}
\affil{University of Exeter, Streatham Campus, Exeter EX4 4QF, United Kingdom}
\author{Claire R. Pollard\footnote{Software Developer and Digital Marketing Coordinator},
             Henry Bucklow\footnote{Product Manager. AIAA Member.}, and
             Mark Gammon\footnote{Technical Director and CADfix Product Manager. AIAA Member.}}
\affil{ITI -- International TechneGroup Limited, Cambridge, United Kingdom}

\begin{document}

\maketitle

\begin{abstract}
We describe a semi-structured method for the generation of high-order
hybrid meshes suited for the simulation of high Reynolds number
flows. This is achieved through the use of highly stretched elements
in the viscous boundary layers near the wall surfaces.  CADfix is used
to first repair any possible defects in the CAD geometry and then
generate a medial object based decomposition of the domain that wraps
the wall boundaries with partitions suitable for the generation of
either prismatic or hexahedral elements. The latter is a novel
distinctive feature of the method that permits to obtain well-shaped
hexahedral meshes at corners or junctions in the boundary layer. The
medial object approach allows greater control on the ``thickness'' of
the boundary-layer mesh than is generally achievable with advancing
layer techniques. CADfix subsequently generates a hybrid
straight-sided mesh of prismatic and hexahedral elements in the
near-field region modelling the boundary layer, and tetrahedral
elements in the far-field region covering the rest of the domain. The
mesh in the near-field region provides a framework that facilitates
the generation, via an isoparametric technique, of layers of highly
stretched elements with a distribution of points in the direction
normal to the wall tailored to efficiently and accurately capture the
flow in the boundary layer.  The final step is the generation of a
high-order mesh using NekMesh, a high-order mesh generator within the
Nektar++ framework. NekMesh uses the CADfix API as a geometry engine
that handles all the geometrical queries to the CAD geometry required
during the high-order mesh generation process. We will describe in
some detail the methodology using a simple geometry, a NACA wing tip,
for illustrative purposes. Finally, we will present two examples of
application to reasonably complex geometries proposed by NASA as CFD
validation cases: the Common Research Model and the Rotor 67.
\end{abstract}

\newpage

\section*{Nomenclature}
{\renewcommand\arraystretch{1.0}
\noindent\begin{longtable*}{@{}l @{\quad=\quad} l@{}}
1D, 2D, 3D                                   & One-, two-, and three-dimensional, respectively \\
API                                               & Application programming interface \\
BFGS                                           & Broyden--Fletcher--Goldfarb--Shanno optimization algorithm \\
CFD                                             & Computational fluid dynamics \\
CFI                                               & CADfix interface: its model API \\
CRM                                             & Common research model \\
STEP                                             & Standard for the exchange of product model data \\
$\Omega$                                   & High-order element \\
$\Omega_\textrm{st}$                  & Reference element \\
$\widetilde{\Omega}_\textrm{st}$ & Subelement of the reference element \\
$\xi$                                              & Parametric coordinates in the reference element \\
$\chi$                                          & Mapping from reference element to high-order element \\
$f$                                              & Affine mapping from reference element to subelement \\
$J_f(\xi)$                                    & Determinant of the Jacobian of the mapping $f$ \\
$P$                                            & Polynomial order \\
$r$                                             & Growth ratio of boundary-layer element heights in the direction of the normal.
\end{longtable*}
}

\section{Introduction}
\label{intro}
\lettrine{O}{ne} of the bottlenecks to the development of high-order
CFD simulations of high-Reynolds number flows and their industrial uptake
is mesh generation \cite{Vincent+Jameson-2011,Wang-2013}. The main 
challenge is to systematically and robustly
generate valid curvilinear high-order boundary-conforming meshes which 
incorporate stretched elements in the near-wall boundary-layer regions. 
If the complexity of the computational domain lends itself to structured multi-block 
decomposition \cite{Armstrong-2015}, then the mapping between the blocks and 
the unit cube provided by this approach facilitates high-order and boundary-layer 
meshing, but domain decomposition for general domains remains
a very difficult and open problem.

Current unstructured high-order mesh generators are based on \textit{a posteriori} 
approaches that deform a coarse linear mesh to accommodate the curvature at the
boundary, see for instance a brief review of these methods in reference \cite{turner-2017a}. 
Robust mesh generators are available for generating the linear mesh, but \textit{a posteriori} 
high-order mesh generators of curvilinear meshes tend to have difficulties in ensuring the 
validity of the mesh when highly stretched elements typical of boundary-layer meshes are present.

Here we propose a semi-structured approach that combines two complementary mesh generation
procedures. We employ a linear mesh generator within the commercial software CADfix \cite{CADfix}, 
that employs the medial object approach to decompose the domain into partitions which can be 
discretized into structured or unstructured meshes. A restriction of such partitioning to the near-wall
regions and an appropriate design of the medial object partitioning reduces significantly the 
complexity of the generation process and makes it possible to obtain high-quality boundary-layer 
type hybrid meshes near the wall surfaces. Further, CADfix provides powerful CAD healing and
modification tools through an interface, CFI, for handling CAD geometrical
operations and queries. The high-order mesh is generated by the open-source 
code NekMesh which is part of the Nektar++ spectral/\textit{hp} element framework 
\cite{cantwell-2015}. All the geometrical interrogations to the CAD definition of the 
boundary of the computational domain are handled by NekMesh via CFI that provides
direct access to the data structures describing the CAD geometry and the linear mesh. 

The generation of a high-order mesh using this semi-structured approach involves
two steps.  We first generate a straight-sided mesh using CADfix with a coarse
boundary-layer mesh composed of a single layer within the medial object based
partitions adjacent to the wall boundaries. Additional points are then added,
following essentially the method described in reference
\cite{Sherwin+Peiro-2002}, to obtain a high-order curvilinear mesh compliant
with the CAD definition.  Next, a boundary-layer mesh is generated using the
isoparametric approach proposed in reference \cite{moxey-2015a} where elements
in the coarse mesh adjacent to the wall are subdivided along the normal
direction according to a user-defined resolution. In this work, we outline an
extension to this method that leverages the medial-object decomposition
available through CFI to generate high-quality meshes in corners and junctions
by performing splitting in two separate directions perpendicular to each surface
of the corner section. This approach is very flexible, modular, and permits
defining a variety of resolutions from a base coarse high-order mesh that
remains unchanged.

These procedures will be described in more detail in the following
sections. Section \ref{sec:CAD} describes the CAD interface that
handles the geometrical queries during the generation of both linear
and high-order meshes. Section \ref{sec:linear} provides an overview
of the medial object approach and discusses its application to the
decomposition of the domain into near-field and far-field regions
which are then subdivided into blocks that are discretized into a
hybrid linear mesh.  The generation of a high-order mesh from this
linear mesh is described in Section \ref{sec:high_order}. Finally, the
methodology is applied to the generation of high-order meshes for
two geometries and the meshes presented in
Section \ref{sec:examples}.

\section{CAD interface for geometrical queries}
\label{sec:CAD}
Processes for both linear and high-order meshing regularly interrogate
the CAD geometry and thus a robust CAD interaction is
required. NekMesh provides a lightweight wrapper that hides the
complexity and size of the CAD interface from users and developers. In
the examples presented here we have used CFI, the CAD interface of
CADfix \cite{CADfix}, but NekMesh also provides a CAD back-end to 
OpenCascade~\cite{OpenCascade} as its CAD engine.

The use of CADfix, and its interface CFI, is motivated by the more
stringent requirements on CAD quality for high-order meshing. CAD
representations that may work very well within linear mesh generators,
may not work for their high-order counterpart. For example, distortion
levels in the surfaces, which might be perfectly acceptable for
generating linear meshes, could induce poor quality or invalid
elements in high-order meshes. Therefore access to high quality CAD
and CAD repair tools for poor quality CAD, along with a robust CAD
interface, is vital to the creation of robust quality high-order
meshing tools.

\begin{figure}[htbp!]
  \centering
  \includegraphics[width=0.5\textwidth]{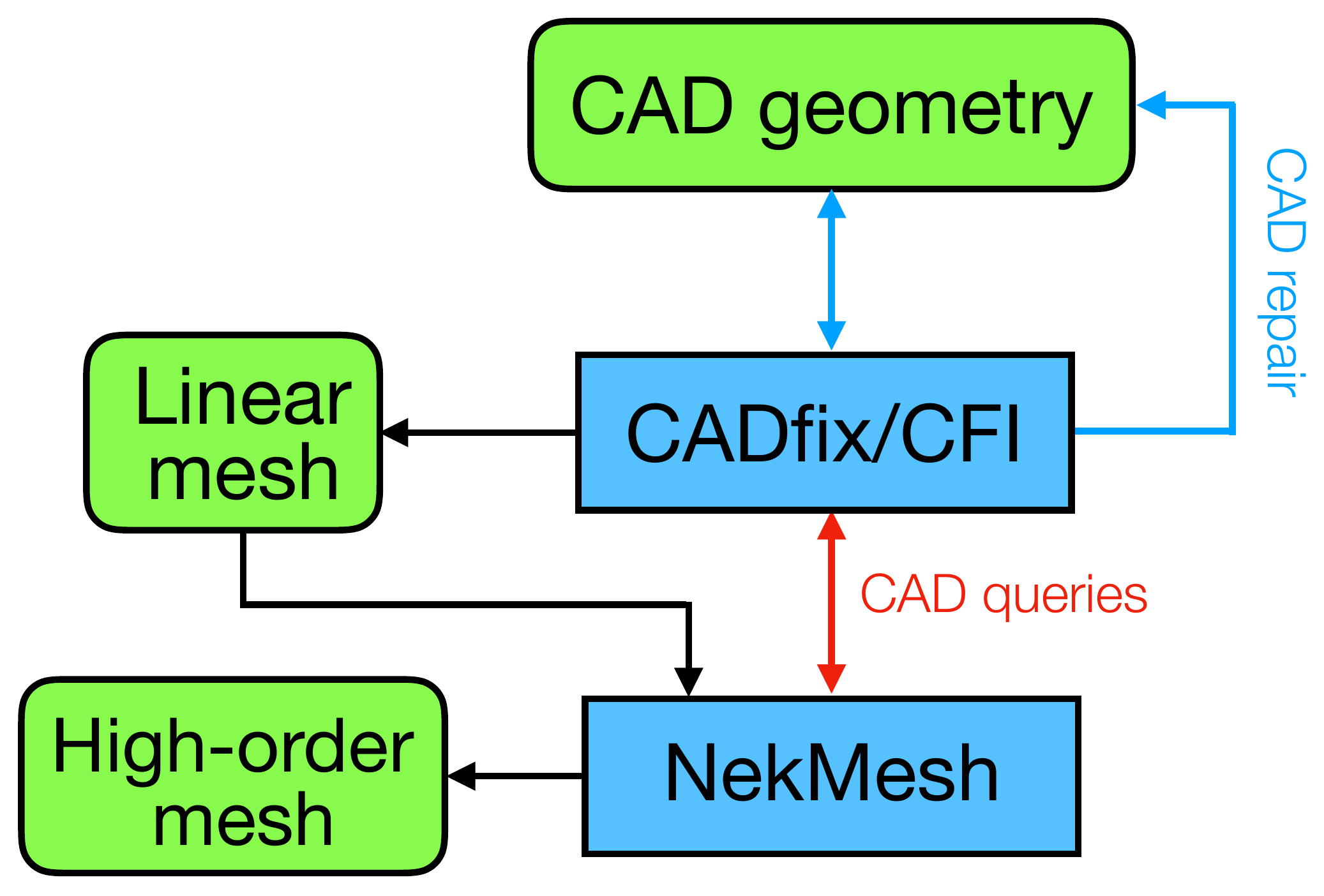}
 \caption{Flowchart of the proposed semi-structured approach to
   high-order mesh generation.  NekMesh interrogates the CAD geometry
   via CFI, the CADfix API. The high-order mesh is produced by NekMesh
   from a linear mesh generated by CADfix.}
\label{fig:pipeline}
\end{figure}

The flowchart of Fig.~\ref{fig:pipeline} depicts the integration of CFI into NekMesh.  
In a nutshell, a CADfix session produces a linear mesh that
NekMesh will read via CFI and process through its own high-order 
routines.  More details of the method will be given in the following sections.

\section{Linear mesh generation via the 3D medial object}
\label{sec:linear}

In order to produce a high-order mesh, we first need to generate a
linear mesh. CADfix is a commercially available tool with
functionality covering the import, preparation and interrogation of
geometry but the 3D medial object based partitioning and linear mesh
generation uses results from active research projects
\cite{Bucklow-2017} and are not yet commercially available. However,
in order to provide appropriate geometry and prismatic meshes for
upgrade to high-order, CADfix has additional functionality which has
been designed for this framework and is under active development.
There are several automatic and semi-automatic tools which are
included in the pipeline for generating linear meshes. First, the
geometry is prepared to repair any CAD defects and to define a valid
domain. Second, we automatically subdivide the domain to create our
partitions for meshing. Finally, the partitions are meshed using a
coarse set of divisions, designed to be balanced, well aligned and to
allow periodicity at the boundary. Each part of this process has been
designed to be suitable for \textit{a posteriori} high-order mesh
generation. In the following sections, these steps are described in
more detail.

To illustrate the various steps of the procedure for constructing a
high-order mesh, we will employ a simple geometrical domain that
consists of an unswept wing of rectangular planform composed of
NACA0012 aerofoil sections and a round tip, essentially a wing tip,
enclosed in a rectangular box.  This geometry is also of aerodynamic
interest as a case study of vortex roll-up proposed and experimentally
measured by Chow et al. \cite{Chow-1997} which has been used in CFD
validation studies, see for instance \cite{Lombard-2016}.

\subsection{Geometry preparation}
Not all CAD models are suitable for CFD analysis. CAD geometry often lacks outer
domain definitions, it may have defects such as sliver surfaces or small gaps 
and the geometry may not be watertight. For our purposes we require a
\textit{CFD-ready CAD geometry}: the fluid domain must be a watertight
CAD solid. CADfix can import the geometry from a wide range of design
sources and provide automatic, manual and diagnostic driven tools for
repairing poor quality CAD geometry, constructing outer domain
boundaries and building a watertight and well connected CAD model. The
3D medial object algorithm also needs a certain level of quality from
the input CAD model. Sharp corners, large vertex-face and edge-face
gaps all need to be repaired before the medial object can be generated
to guide the partitioning and ultimately the meshing. As the domain
partitioning and meshing process respects the CAD topology,
excessively short edges and narrow sliver faces should also be
removed. The requirements outlined here are not that different to
those imposed by standard surface and volume meshing algorithms, and
typically can be automatically detected and removed.

\subsection{3D medial object}
\begin{figure}[htbp!]
	\centering
	\includegraphics[width=0.7\textwidth]{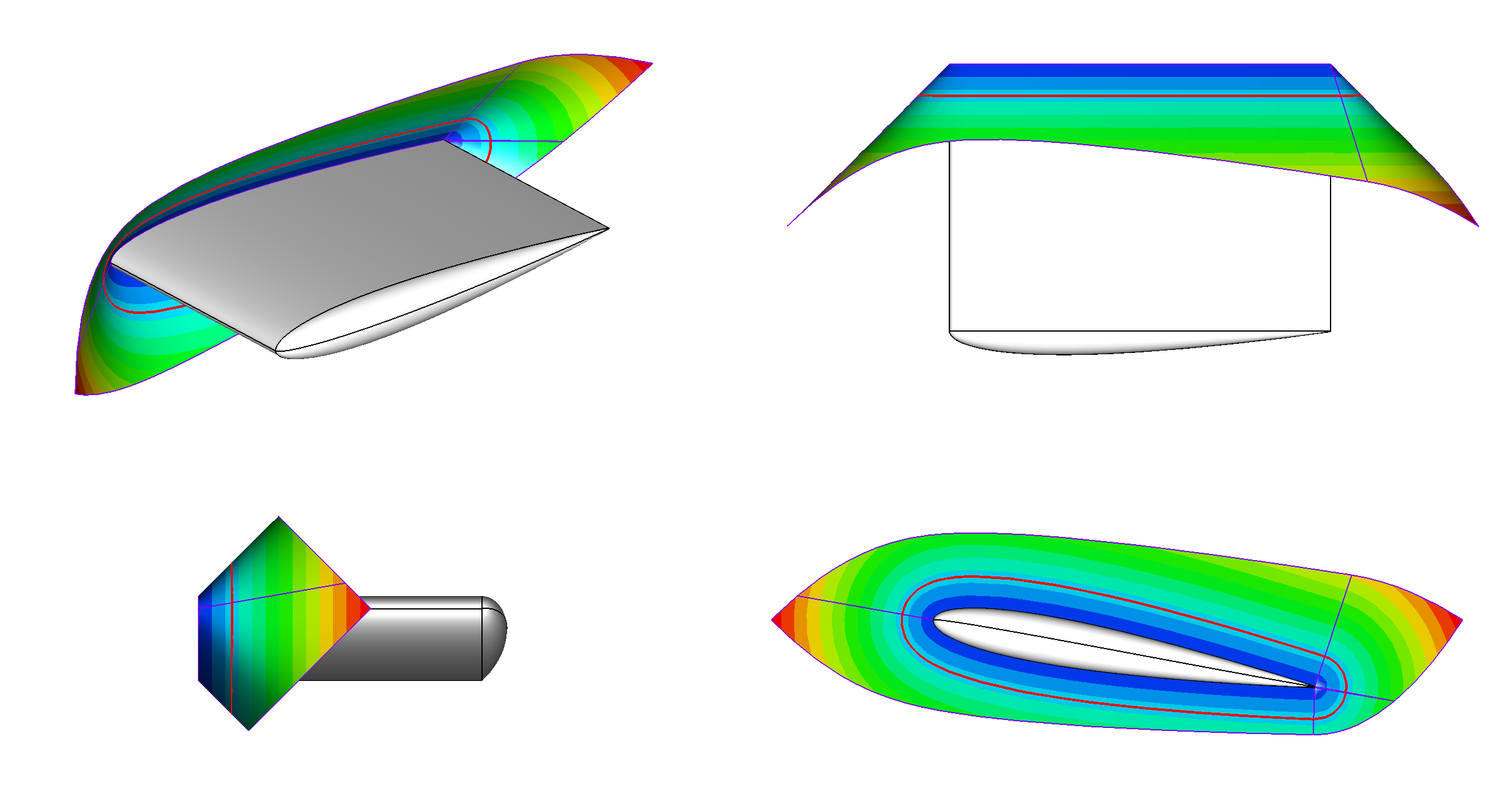} 
	\caption{Example of a 3D medial object for the fluid domain
          about the NACA wing tip.}
	\label{fig:medial}
\end{figure}

The medial axis, first introduced by Blum~\cite{Blum-1967}, is a
method for analysing shapes. For a fluid domain, it can be defined as
the set of all points in the domain which have more than one closest
point on the boundary of the domain. If these points are taken
together with their distance to the domain boundary (the
\textit{medial radius}), they form a complete description of the flow
domain. The medial axis is computed and returned as a non-manifold CAD
object called the medial object, which contains extra information to
describe the relationships between the different components of the
medial object along with medial radius information. See
Fig.~\ref{fig:medial} for an example 3D medial object of the fluid
domain about the NACA wing tip.

The medial object can be used for structured meshing, feature
recognition and mid-surfacing as well as the automatic partitioning
used here, and robust generation of the medial object has been a long
standing challenge for the CAE community. Our algorithm is based on a
domain Delaunay triangulation~\cite{Sheehy-1994}, and recent
developments~\cite{Bucklow-2014} allow it to robustly work on a range
of production CAD models or in the air volume around such models.

\subsection{Using the 3D medial object for partitioning}

\begin{figure}[htbp!]
	\centering
	\includegraphics[width=0.4\textwidth]{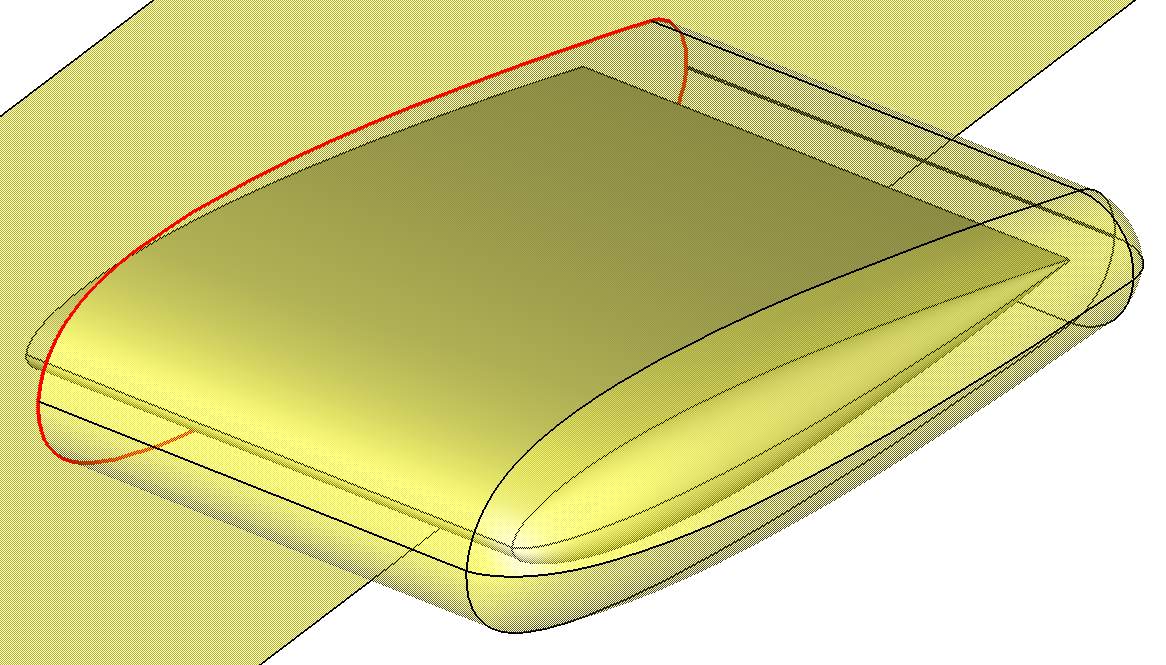}
	\caption{A "shell" around the NACA wing tip geometry that
          divides the domain into two partitions: near-field (close to
          body) and far-field (away from body). The near-field
          partition will be used to generate a boundary-layer mixed
          mesh of prisms and hexahedra and a tetrahedral mesh is
          generated in the far-field partition.}
	\label{fig:shell}
\end{figure}

The 3D medial object is used to guide our partition generation in
complex junctions. Firstly, we must construct the 3D medial object,
and then use this to generate an offset surface, or \textit{shell},
from the boundaries of the fluid domain. The medial object is used to
locate lines where simply offsetting the CAD faces would cause the
shell to self-intersect, known as \textit{medial halos}
(the red lines in Fig.\ref{fig:medial}). The shell (Fig.~\ref{fig:shell}) generated
splits the fluid domain into two partitions: one near-field partition
close to the boundary and one far-field partition. The near-field
partition is subdivided into multiple smaller partitions using feature
lines on the CAD model to guide the location of the partition faces.

If the fluid domain contains a sharp concave corner or edge (for
example, at a wing/fuselage junction), flows will occur with
potentially large velocity gradients in two or three
directions. Ideally our coarse linear mesh requires elements aligned
with these principal directions. Using the medial object, there are
several options available to achieve a mesh suitable for high-order
upgrade.

\begin{figure}[htbp!]
	\centering
	\begin{tabular}{ccc}
		\includegraphics[width=0.3\textwidth]{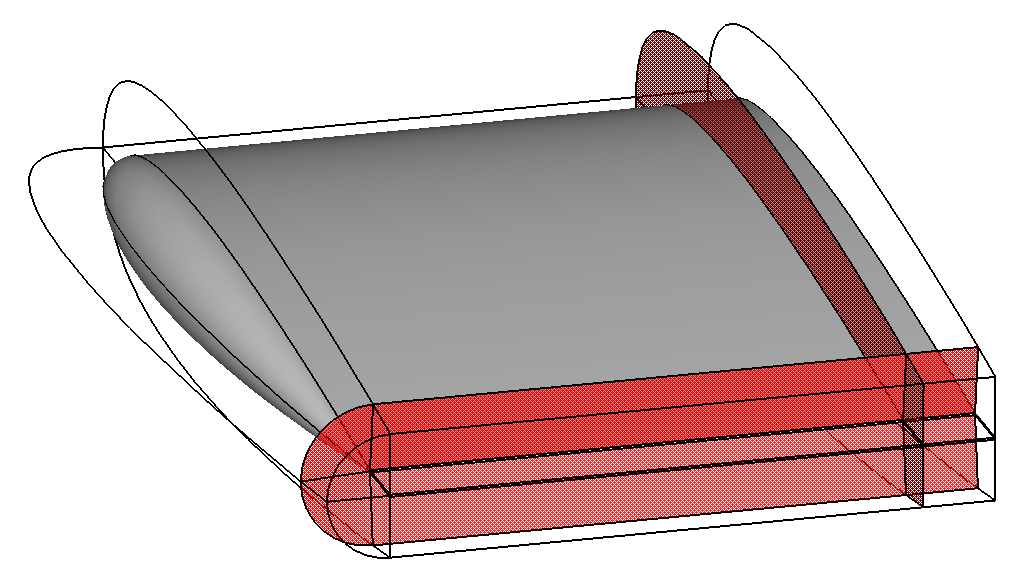}  &
		\includegraphics[width=0.3\textwidth]{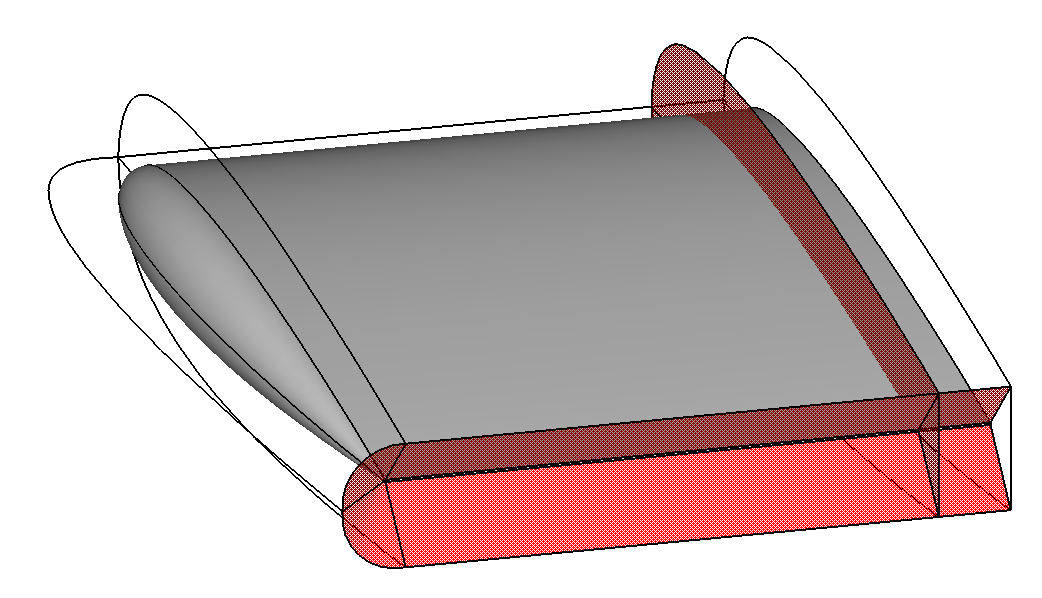}  &
		\includegraphics[width=0.3\textwidth]{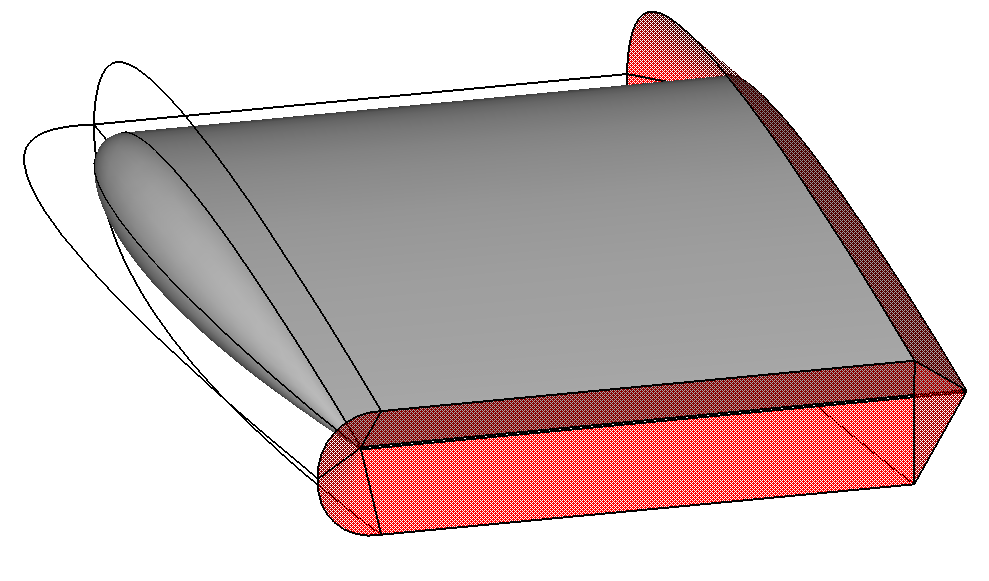} \\
		(a) & (b) & (c) 
	\end{tabular}
	\caption{An illustration of the different shell structures designed to create:
		(a) H-type, (b) C-type, and  (c) O-type topologies. The C-type topology has been adopted 
		here to generate a prismatic boundary-layer linear mesh.}
	\label{fig:topology}
\end{figure}

The medial halos and medial object itself can be used to guide
partition construction around concavities in wing root junctions,
giving better mesh alignment when using hexahedral meshing. This gives
us an H-type topology, similar to those constructed with a structured
multiblock system, as shown in Fig.~\ref{fig:topology}(a). This H-type
topology is not ideal for the highly stretched meshes required for
high-order meshing. Instead, we have chosen a C-type topology,
illustrated in Fig.~\ref{fig:topology}(b), which removes the need for
hexahedral elements along the trailing edge, replacing them with large
prismatic elements. However, this topology still requires hexahedral
elements in the wing root junction partition to allow us to complete
meshing of this partition structure. Therefore an O-type topology
shown in Fig.~\ref{fig:topology}(c) has been designed to generate the
highly stretched meshes needed to simulate near-wall flows. This
allows a structured prism dominant linear mesh to be built in all
near-field blocks, removing all hexahedral elements in junction
regions. The C-type topology has been adopted in the following examples 
to generate coarse linear boundary-layer meshes.

\begin{figure}
   \centering
     \includegraphics[width=0.6\textwidth]{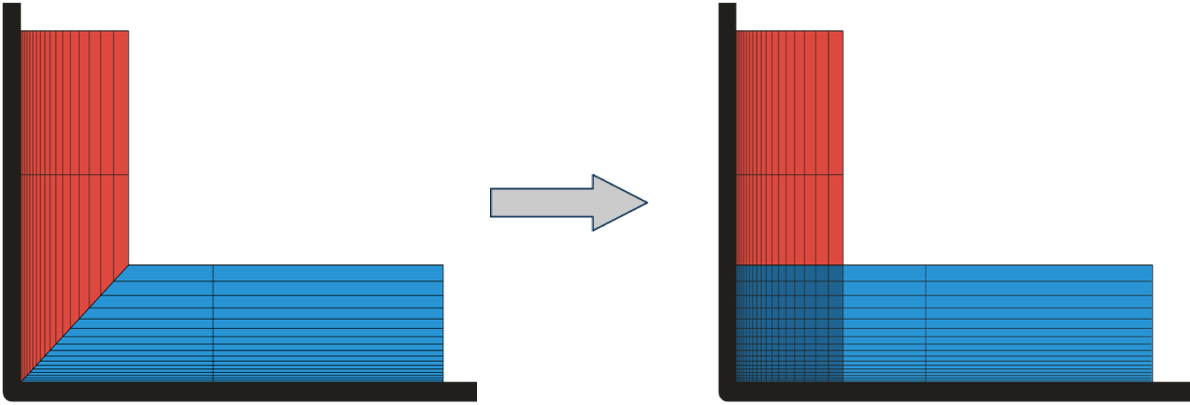}   
   \caption{C-type mesh advantages: the medial object guides partitioning in junction regions, 
   avoiding the effects of layer stopping at corners.}
   \label{fig:rotor67-1}
\end{figure}

The use of a C-type topology, together with hexahedral elements, enhances the quality of the linear mesh at
junctions. This is illustrated by Fig.~\ref{fig:rotor67-1} which compares the meshes obtained 
with a O-type and a C-type approach and shows that the C-type mesh avoids
the distortion of prismatic layers at the corners. The O-type mesh could potentially prevent the propagation 
of the prismatic layers due to self-intersection.
  
\subsection{Linear mesh generation}

The medial object based partitioning of the flow domain has been
designed for use with certain mesh styles.

The O-type near field topology allows the use of prismatic linear
elements which can be swept from the CAD surfaces through the
near-field partition to interface with a tetrahedral mesh in the far
field, with only one element generated through the thickness of the
near field partition.

We follow a bottom-up mesh generation process to ensure the mesh is
fully conformal between all partitions. We mesh lines, then surfaces
are meshed with elements conforming to the lines, and finally the
volumes of the partitions are meshed with elements which conform to
the faces. The C-type topology features a structured hexahedral
junction partition, and in this case the line meshes must be
``balanced'' to satisfy rules which are imposed via a structured mesh
style. This is solved as an integer programming problem~\cite{Tam-1993}, 
and solved using an open source solver~\cite{lpsolve-2017}. 
To further ensure good quality in the final mesh,
a least-squares optimisation is performed to the line nodes to reduce
potential skew.

\begin{figure}[htbp!]
   \centering
     \includegraphics[width=0.5\textwidth]{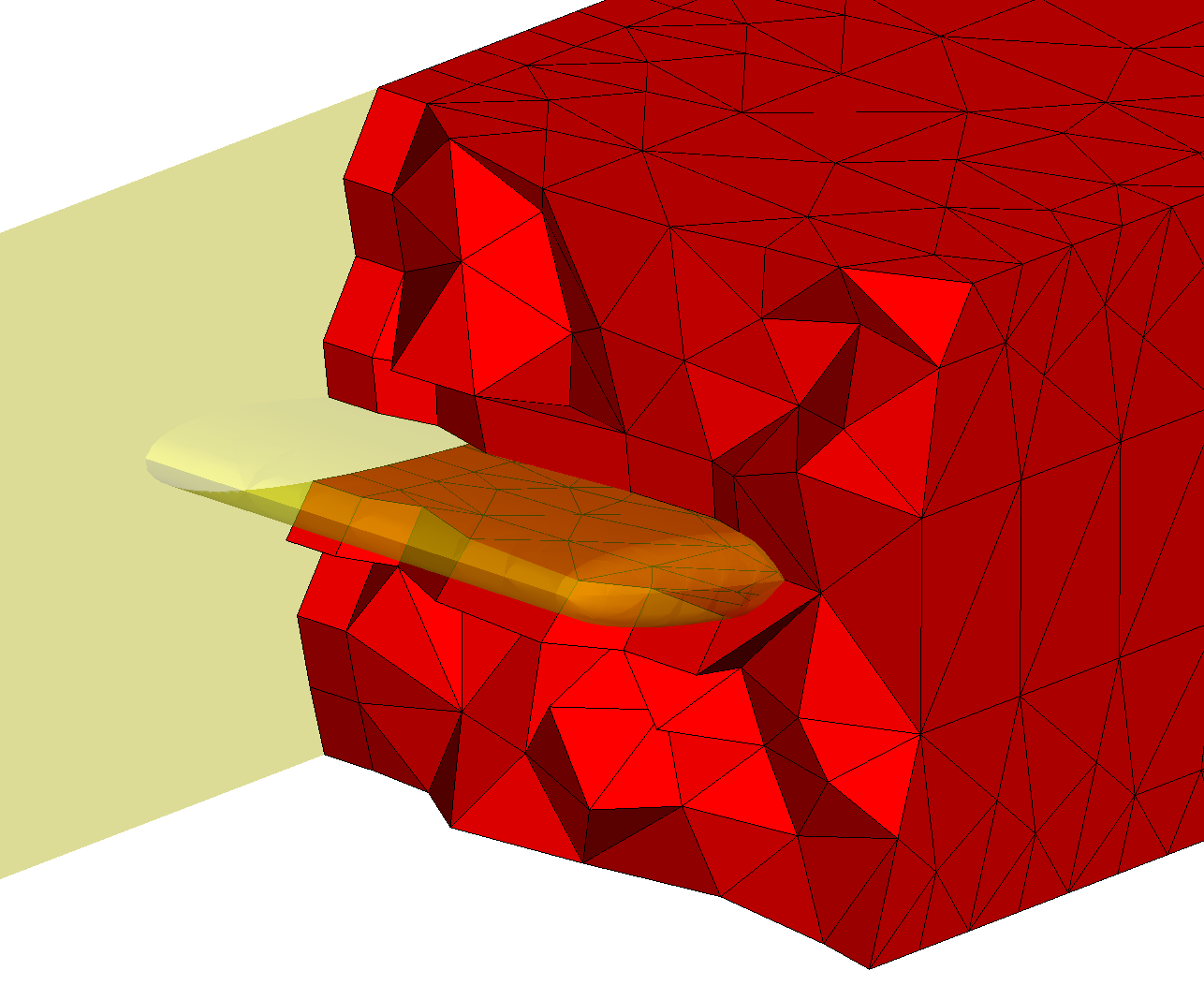}   
   \caption{NACA wing tip: view of the linear mesh.}
   \label{fig:NACA-linear}
\end{figure}

The swept meshing of the partitions is performed by Delaunay
triangulations of the designated template faces. The surface meshes
have a simple 30 degree turn angle sizing applied to take into account
changes in curvature to ensure the mesh generated is coarse
enough. This Delaunay mesh is swept into prismatic elements using the
CADfix sweep mesher. Once the partition mesh and the tetrahedral
far-field mesh have been completed, a mesh quality test is performed to
make sure all elements produced during the linear meshing stage are
not inverted. An example of a coarse linear mesh obtained with this method
for the NACA wing tip geometry is shown in Fig.~\ref{fig:NACA-linear}.
The boundary-layer mesh in the near-field region consists of 
1\ 224 triangular prisms and 25 hexahedra and the far-field region is 
discretized into 12\ 576 tetrahedra.

\subsection{Linear mesh periodicity}
For the Rotor 67 example that follows in Section~\ref{sec:examples}, a
rotationally periodic mesh is required. The far-field boundary edges and
surfaces need matching divisions so identical meshes can be made to ensure
periodicity of the solution. This is achieved by performing an additional step
in the linear mesh generation process during the balancing and alignment
step. By calculating a rotational transformation which takes one side of the far
field to the other, we can geometrically match edges and copy the divisions from
one side of the outer far field faces to the other. As these edges and surfaces
have already been balanced and aligned we are safe to duplicate the divisions on
the other side of the far field, maintaining the density and quality required
for our coarse meshes.  Once the divisions have been duplicated the volume can
be meshed and quality checked as outlined above.

\section{High-order meshing}
\label{sec:high_order}
The \textit{a posteriori} generation of a high-order mesh from a linear
mesh proceeds in a bottom-up fashion following the ideas proposed in
reference \cite{Sherwin+Peiro-2002}. The additional points required
for the high-order polynomial discretization are incorporated
sequentially first along the curves, then on the surfaces of the CAD
geometry and, finally, in the interior of the domain. The generation
of points along the curves is essentially the one proposed in
reference \cite{Sherwin+Peiro-2002}, the following sections describe
the improvements that we have incorporated into the methodology to
achieve the type of meshes sought in this work.
 
\subsection{Surface optimisation}
Inaccuracies in the representation of the geometry of the
boundary of the computational domain due to CAD distortion, even if small,
could have a significant impact on the accuracy of the flow solution.
To overcome this problem, NekMesh optimises the location of the high-order 
nodes in the mesh to reduce distortion by modelling the mesh entities as spring
networks and minimising their deformation energy. The optimal location of the 
mesh nodes is obtained in a bottom-up fashion. 

The first step is to optimize the location of mesh nodes belonging to edges that lie 
on curves by minimizing the deformation energy of a spring system in the parametric 
space of the curve with the vertices in the linear mesh fixed.
This is followed by the processing of mesh nodes on edges that lie
on the CAD surfaces. Again, we work on the 2D parametric space of the
surface and find the optimal position of the mesh nodes of an edge by minimizing 
the deformation energy of the 3D high-order edge. As a result, the optimised
high-order edge will lie approximately on the geodesic between the two
end points on the surface.
The final step is the relocation of the mesh nodes in interior triangle faces that lie on CAD
surfaces. Here we follow a slightly different approach to the previous ones. The mesh nodes
on the edges of the triangles are fixed and the interior nodes are free to move. Each of the 
free interior nodes is connected to a system of six
surrounding nodes by springs. The minimum deformation energy of this system 
of spring is found using a bounded version of the BFGS algorithm that accounts for the
limits of the parameter space in the CAD entities~\cite{Byrd-1995}. The gradients required 
by the optimization procedure can be evaluated from the CAD information provided by
CFI. This procedure leads to an overall distribution of points which ensures the surface mesh 
is smooth, unless pathological distortion is present in the CAD geometry. 

\subsection{Boundary layer meshing}
The generation of highly stretched elements with high aspect ratios, say 100:1,
which are required to accurately simulate the high shear of boundary-layer flows
at aeronautically-relevant Reynolds numbers, poses a significant challenge for
high-order mesh generation. If the high-order mesh is produced using \textit{a
  posteriori} methods, then curving thin elements in the boundary-layer mesh
will almost certainly produce self-intersecting elements in regions of high
curvature.  To avoid this, we generate high-order boundary-layer meshes by
applying the isoparametric approach \cite{moxey-2015a} to the linear meshes
produced via the medial object.

Firstly, a macro boundary-layer hybrid mesh consisting of a single layer of
hexahedra and triangular prisms is  generated by the medial object method in 
the near-field region and the far-field partition is discretized into tetrahedra.   
The medial object allows us to control the thickness of the near-field region and
the height of the elements to a much greater extent than 
most commercial mesh generators. By selecting a thickness of the shell that
gives enough room to accommodate the surface curvature we reduce the 
likelihood of generating invalid high-order elements within the macro boundary-layer
mesh.

\begin{figure}
  \centering
     \includegraphics[width=0.98\textwidth]{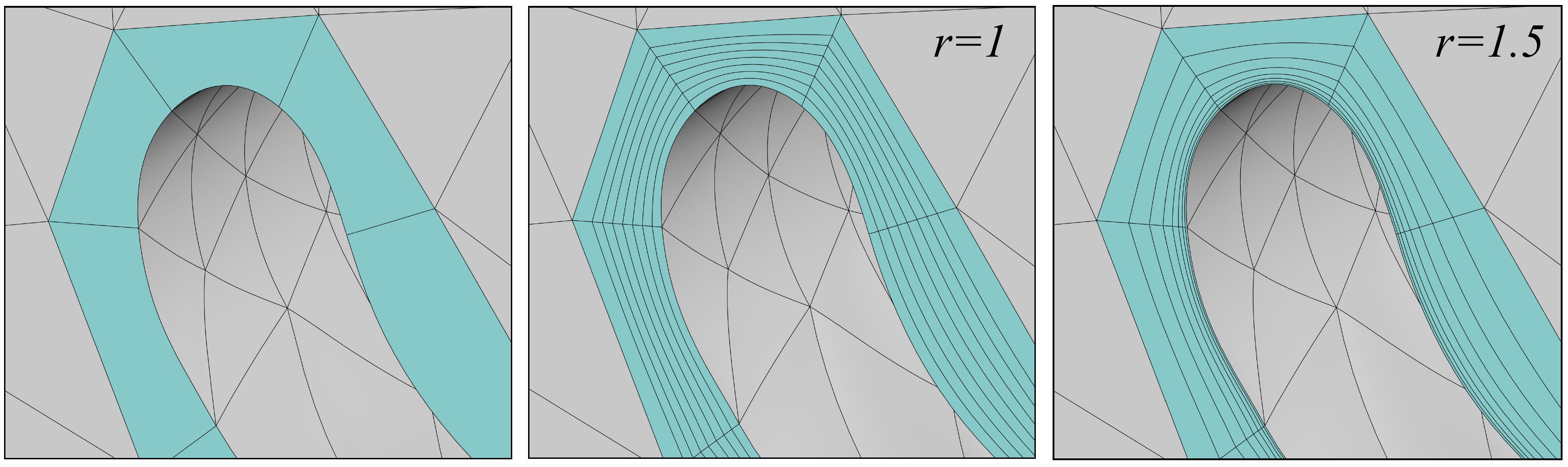}   
     \caption{In the isoparametric approach a macro high-order prismatic element 
     in the near wall region (left figure) is split along the normal to produce elements 
     of high aspect ratio. We prescribe the number of elements
     and the growth ratio of heights between adjacent boundary-layer elements, $r$. 
     For example, $r=1$ (middle figure) corresponds to constant height, and $r=1.5$, 
     increases the element height by half in the normal direction (right figure).}
  \label{fig:bl_mesh}
\end{figure}

The volume generation proceeds next to split these high-order elements using the
isoparametric approach \cite{moxey-2015a}. If the elements are valid, there
exists a bijective mapping $\chi$ between a reference element
$\Omega_{\text{st}}$ and the physical space element $\Omega$. We use the mapping
to introduce subdivisions, according to a user-defined criterion, of the
reference element along the height to generate layers along the normal in the
physical space, as shown in the left-hand side of Figure~\ref{fig:bl-refined}.
This way we can generate very thin boundary layer elements that are themselves
valid if the mapping satisfies certain restrictions, as shown
in~\cite{moxey-2015a,moxey-2015d}. The splitting strategy used here is to
specify a number of subdivisions, or layers, along the parametric coordinate
representing the wall normal and a growth, or progression, rate for the height
of the elements. This progression rate is characterized by a factor, $r$, that
is the ratio of heights of adjacent elements.  The generation of the
boundary-layer mesh by this approach is illustrated in Fig.~\ref{fig:bl_mesh}
which shows a boundary layer region of macro-prisms split to produce highly
curved valid boundary-layer elements with very high aspect ratios.

\subsection{Extension of isoparametric splitting in multiple directions}
The generation of C-type meshes in junction regions allows for the
generation of structured, hexahedral meshes that avoid the effects of
layer stopping in corners and produce higher-quality meshes. However,
a significant downside to this approach from the high-order
perspective is that it breaks the isoparametric splitting of prismatic
stacks of elements that has been used thus far to produce boundary
layer meshes of arbitrary thickness. In this section, we show how this
method can be adapted to deal with C-type boundary layer refinement in
order to generate valid, curved meshes through a straightforward
extension of the technique.

\begin{figure}[htbp!]
  \begin{center}
    \includegraphics[width=0.7\textwidth]{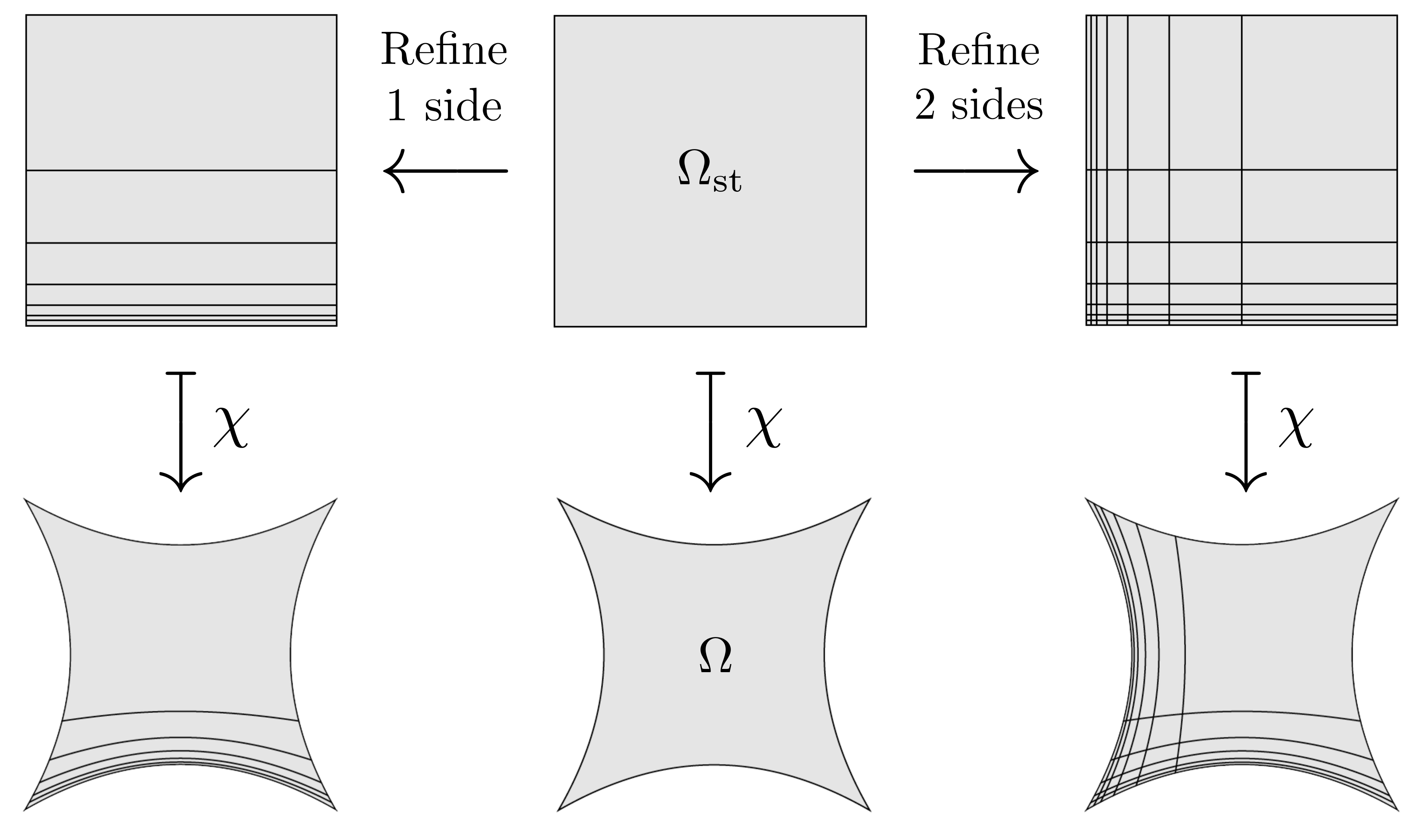}
  \end{center}
  \caption{Adapted version of the isoparametric splitting to work for generating
    high-order junction-adapted meshes.}
  \label{fig:bl-refined}
\end{figure}

The original isoparametric refinement technique first proposed
in reference~\cite{moxey-2015a} splits a valid prismatic macro-element into a
stack of high-order prismatic elements by using the polynomial mapping
that defines the curvature of the element. Mathematically, this
mapping $\chi:\Omega_{\text{st}}\to\Omega$ is defined between a
reference element $\Omega_{\text{st}}$ and a given element
$\Omega$. The key observation in the isoparametric splitting technique
is that to generate the stack of elements, we may split the standard
element $\Omega_{\text{st}}$ into reference sub-elements, and then
apply $\chi$ to these to produce sub-elements in Cartesian space. This
process is depicted visually in Fig.~\ref{fig:bl-refined} for a
simple quadrilateral element. When this procedure is applied to a
boundary layer mesh, the result is a curved valid mesh, regardless 
of the element type.

To apply this technique for the present problem of junction meshing, we require
an adapted version of this approach wherein the reference element is split in
not one but \emph{two} directions that correspond to each surface of the
junction. From the perspective of the mathematical justification for the
validity of the method, this aspect actually makes very little difference. As
noted in reference~\cite{moxey-2015a}, the splitting of the reference element in
the original isoparametric technique can be viewed as an affine mapping
$f:\Omega_{\text{st}}\to\widetilde{\Omega}_{\text{st}}$, where
$\widetilde{\Omega}_{\text{st}}$ is a sub-element of $\Omega_{\text{st}}$. The
curvature mapping of a sub-element of the Cartesian element $\widetilde{\Omega}$
can then be viewed as the composition
$f\circ \chi : \Omega_{\text{st}}\to\widetilde{\Omega}$.  Then, as long as $f$
is defined such that its Jacobian determinant $J_f(\xi) > 0$ for all
$\xi\in\Omega_{\text{st}}$, then this new mapping is valid so long as $\chi$ is
also valid.

To adapt this technique for our junction meshing problem, we therefore
require a slightly different refinement strategy, as depicted in
Fig.~\ref{fig:bl-refined}, where the standard element is split in
each direction. A small extension to the method has been applied which
computes the orientation of the hexahedron and alters the distribution
of the splitting points in the reference element accordingly, but the
core of the method remains mostly the same.

\begin{figure}[htbp!]
  \begin{center}
  \begin{tabular}{ccc}
    \includegraphics[width=0.45\textwidth]{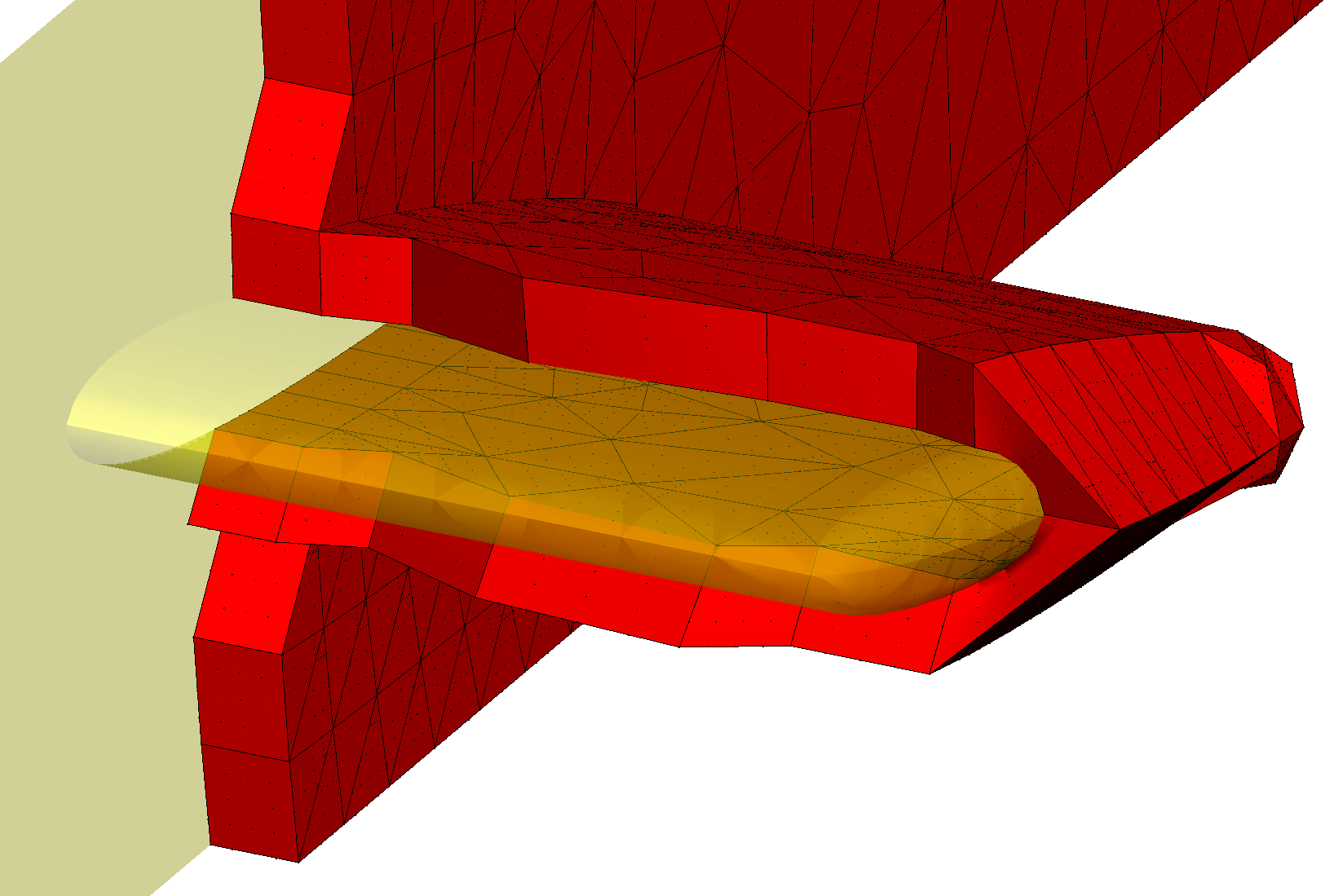} &  &
      \includegraphics[width=0.45\textwidth]{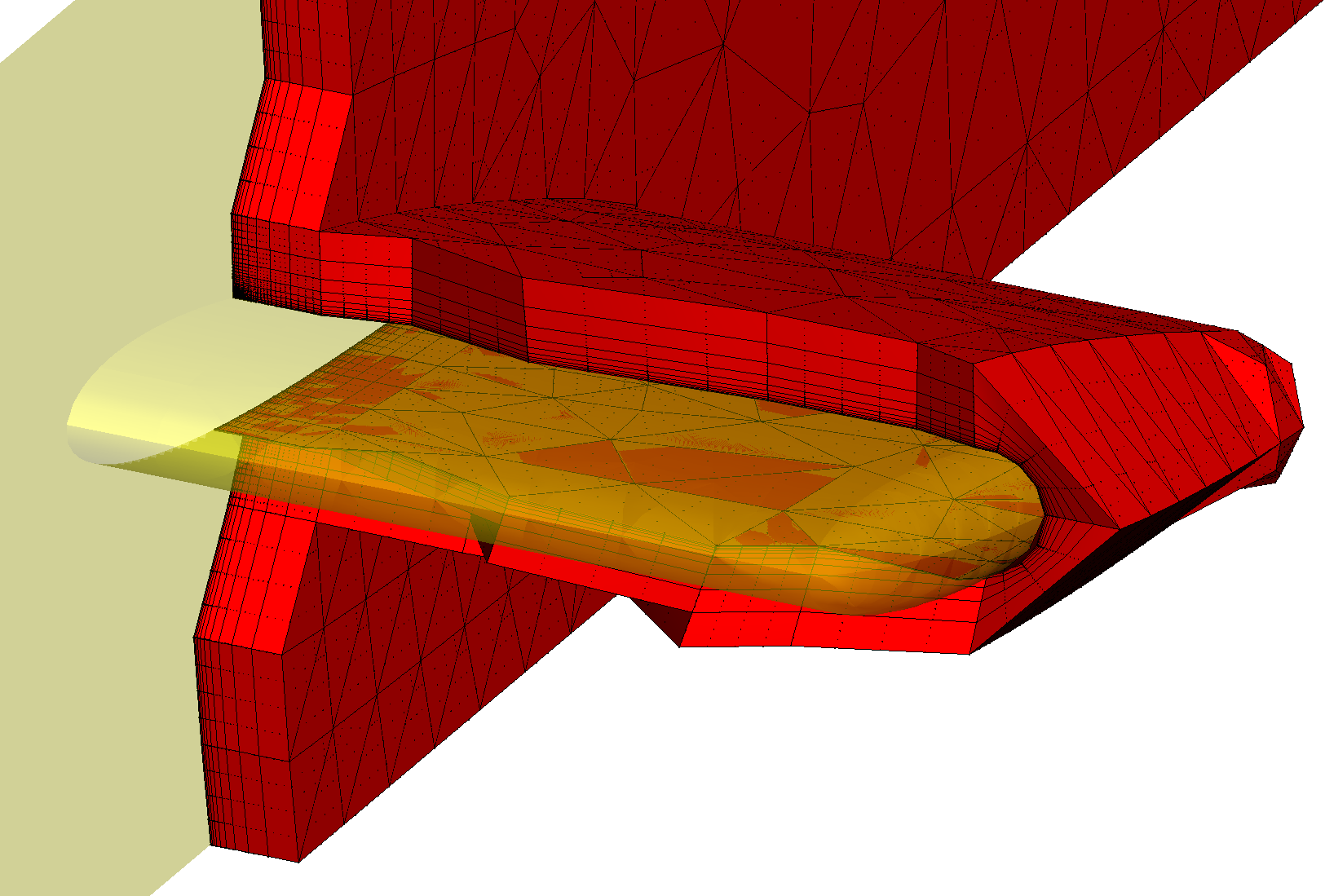} \\
      (a) & & (b) \\
            & & \\
          \includegraphics[width=0.42\textwidth]{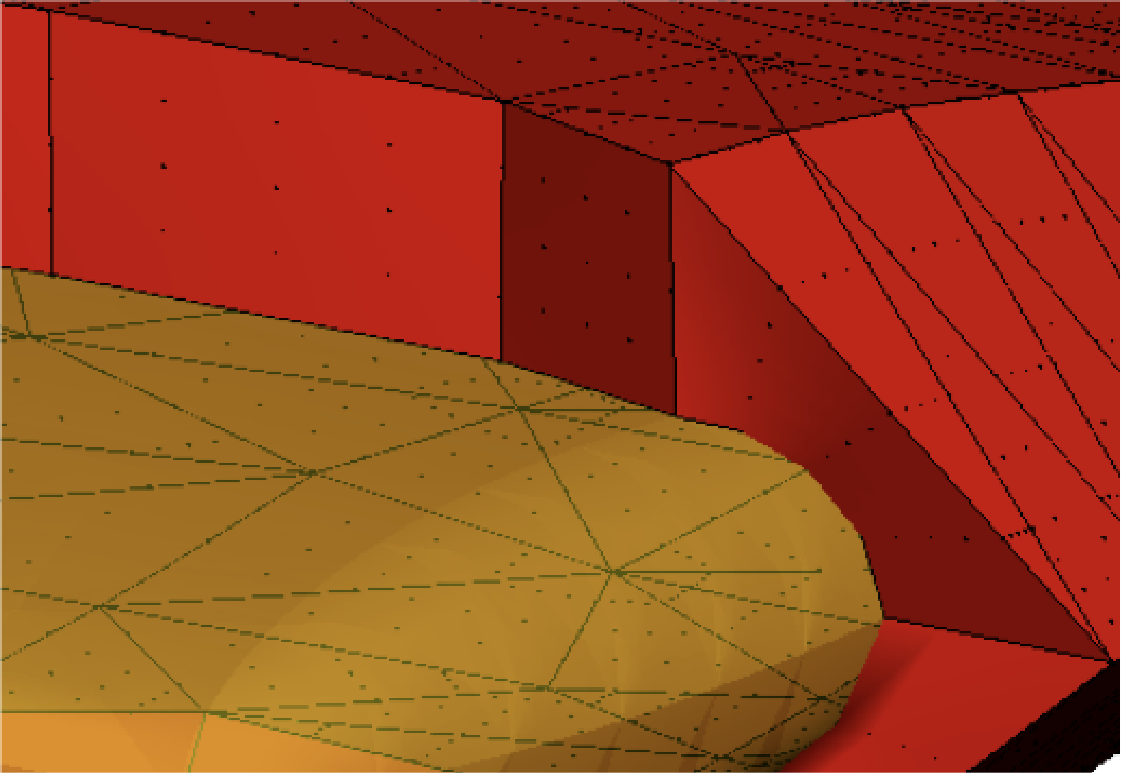} & &
      \includegraphics[width=0.42\textwidth]{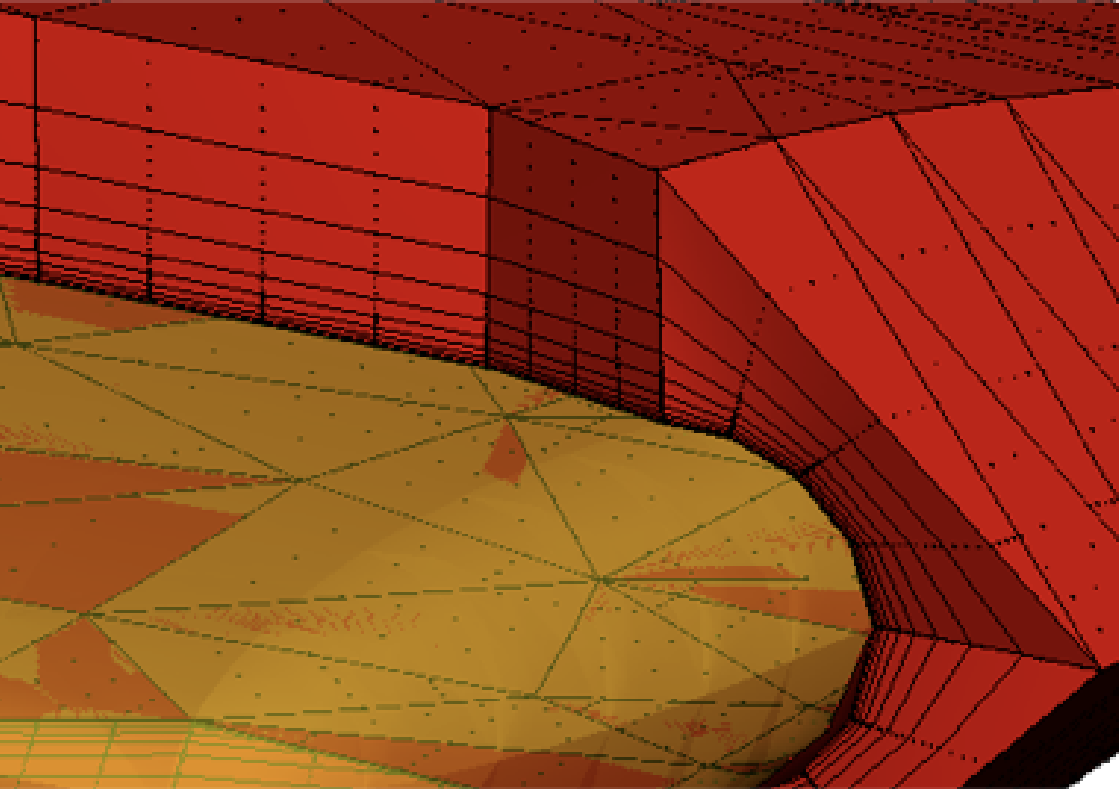} \\
     (c) & & (d)
   \end{tabular}
  \end{center}
  \caption{NACA wing tip boundary-layer mesh: (a) coarse high-order mesh, and (b) after the isoparametric splitting;
  (c) close-up near the tip of the coarse high-order mesh and (d) after the isoparametric splitting.}
  \label{fig:NACA-bl-mesh}
\end{figure}

An example of the boundary layer mesh generated by this method for the wing tip 
geometry is shown in Fig.~\ref{fig:NACA-bl-mesh}. The linear mesh in the near-field
region has been subdivided into 10 layers using a growth rate, $r=1.5$. The resulting
boundary-layer mesh is formed by 12\ 240 prismatic and 2\ 500 hexahedral elements
with a maximum aspect ratio of 70.

\subsection{Volume meshing}

The introduction of the curvature of the CAD surfaces onto the high-order
surface triangulation produces high-order elements in the interior of
the volume with curved faces and edges which could become invalid. 
Controlling the thickness of the near-field via the medial object permits
the generation of linear boundary-layer meshes that can accommodate
the deformation induced by surface curvature without producing invalid 
elements. The positions of the additional nodes required for the polynomial 
representation of the high-order elements are obtained by means of 
a mapping between a reference element and the physical element 
which accounts for the presence of curvature on its faces and edges 
lying on the CAD definition, whilst the other edges are straight and their 
faces planar.

\begin{figure}[htbp!]
   \centering
     \includegraphics[width=0.5\textwidth]{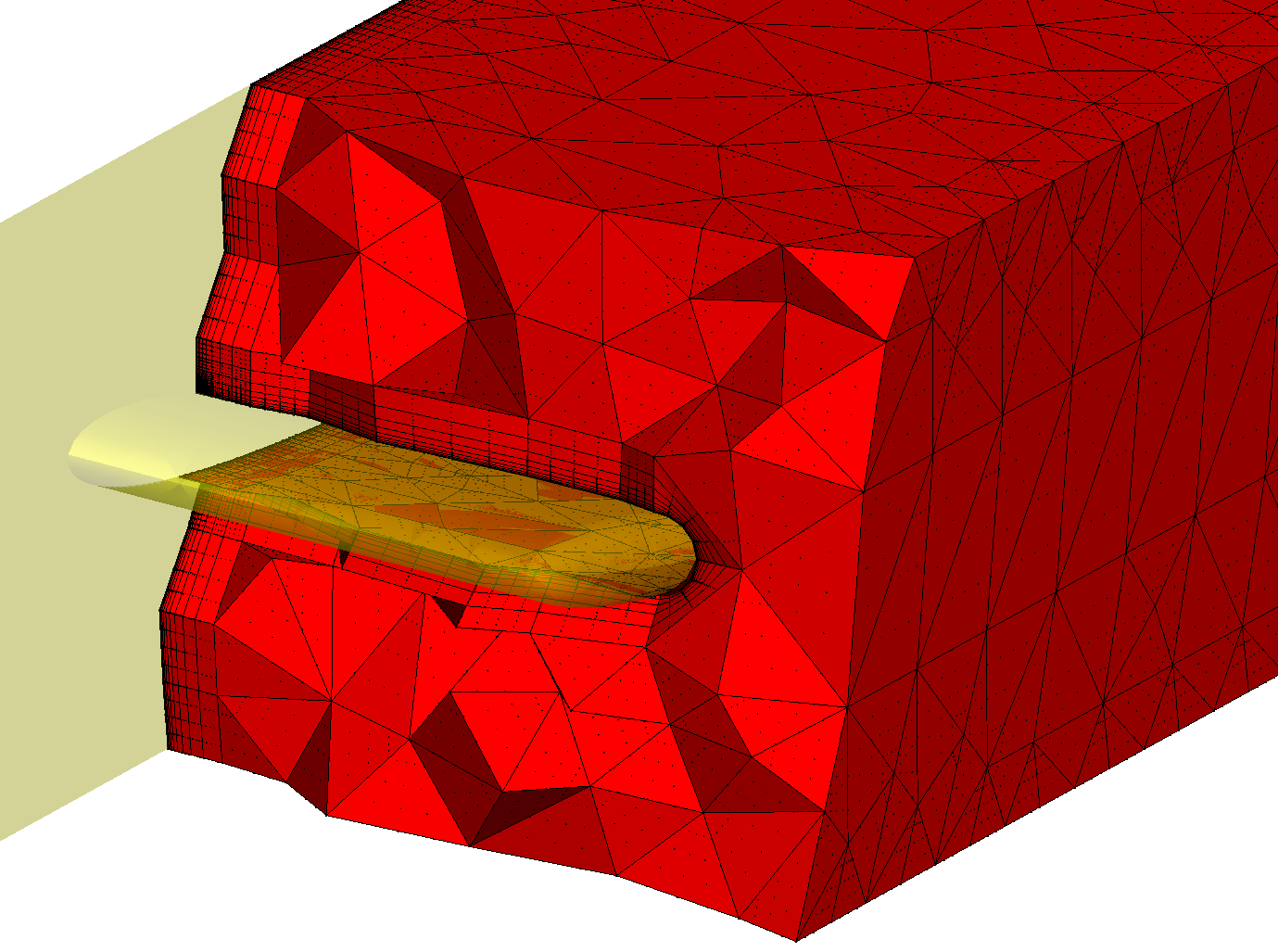}   
   \caption{NACA wing tip: view of the high-order mesh.}
   \label{fig:NACA-high-order}
\end{figure}

A high-order mesh of the volume is shown in Fig.~\ref{fig:NACA-high-order}
which contains, in addition to the elements in the boundary-layer mesh, 
12\ 576 tetrahedral elements of polynomial order 4.

\section{Examples of Application}
\label{sec:examples}
This section presents an illustration of the proposed mesh generation
methodology and the high-order meshes it produces using two geometries
proposed by NASA for CFD validation: the Common Research Model and the
Rotor 67.  These are described in the following sections and, for
reference, all the corresponding high-order meshes have been generated
using a polynomial order of 4.
   
\subsection{NASA Common Research Model}
The Common Research Model (CRM) presented here is one of 
the five configurations designed by NASA \cite{CRM-2008} for CFD validation. 
It is a wing/body alone configuration with a fuselage with a maximum radius of 0.17m, 
and a 35 degrees backward-swept wing of aspect ratio of
9 and span of 1.60m. 

In the first instance, we processed the original definition of the CAD geometry in STEP 
format \cite{CRM-step} through CADfix to clean it and fix a number of
inconsistencies and severe distortions that might prevent the successful generation of the
medial object decomposition and the high-order mesh.

\begin{figure}
\begin{center}
  \begin{tabular}{ccc}
      \includegraphics[width=0.49\textwidth]{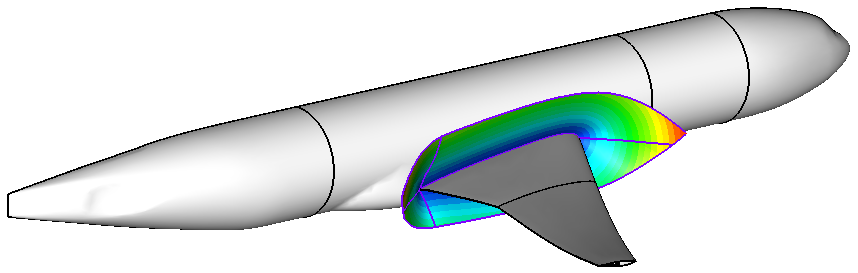} && 
      \includegraphics[width=0.49\textwidth]{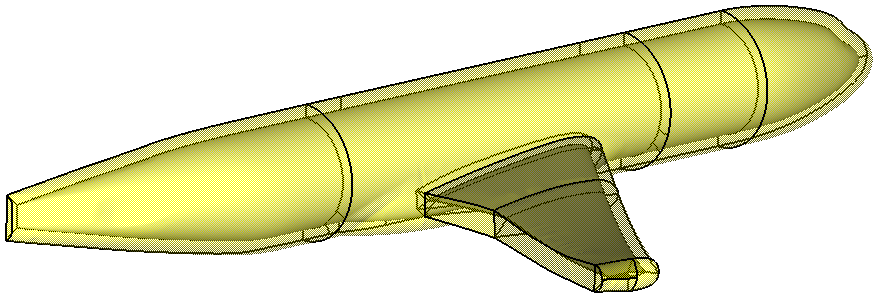} \\
	(a) &  & (b)
 \end{tabular}
	\caption{NASA CRM medial object: (a) interface of the medial
          object at the wing-fuselage junction, and (b) partitions in
          the near-field region.}
	\label{fig:CRM_MO_1}
\end{center}
\end{figure}

\begin{figure}
\begin{center}
  \begin{tabular}{cc}
	\includegraphics[width=0.54\textwidth]{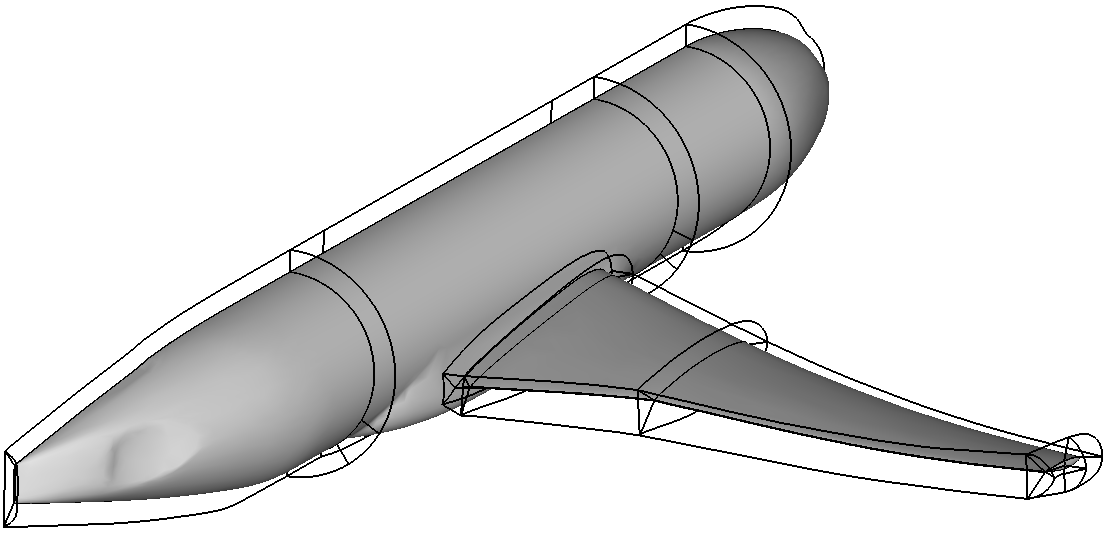} & 
        \includegraphics[width=0.44\textwidth]{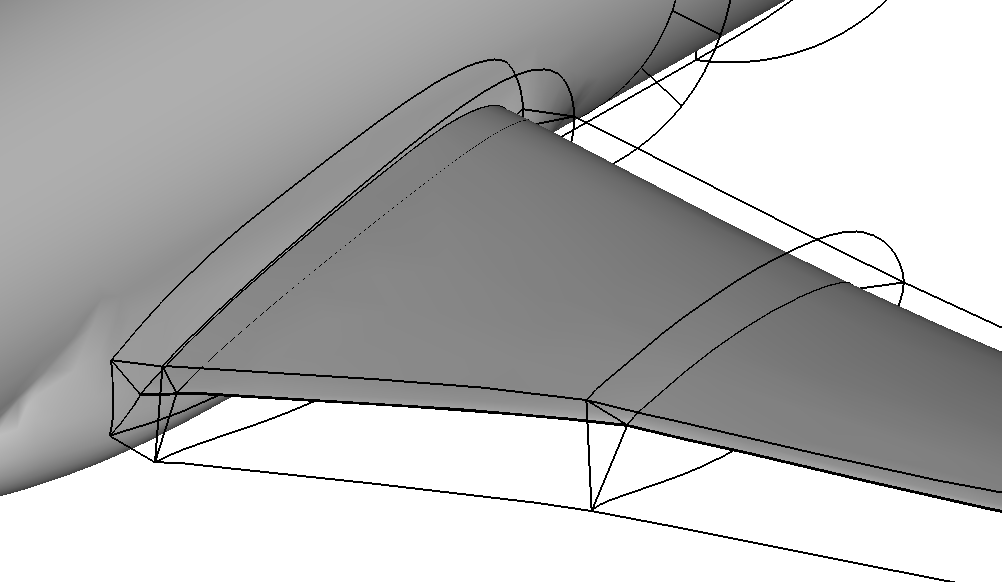} \\
	(a) & (b)
  \end{tabular}
	\caption{A wireframe representing the edges of the partitions
          in the near-field region: (a) global view of blocks around
          wing and fuselage, and (b) close-up near the wing.}
	\label{fig:CRM_MO_2}
\end{center}
\end{figure}

The medial object interface, depicted in Fig.~\ref{fig:CRM_MO_1}(a), was designed to generate a 
hexahedral mesh at the wing-fuselage junction. Fig.~\ref{fig:CRM_MO_1}(b) shows the block partitioning
of the near-field region that provides the framework for generating the boundary-layer mesh around the 
aircraft. Fig.~\ref{fig:CRM_MO_2} provides a more detailed view of the blocks in the near-field region 
via a wireframe representation of the edges of the partitions in that region.

\begin{figure}
	\centering
	\begin{tabular}{cc}
        \includegraphics[width=0.49\textwidth]{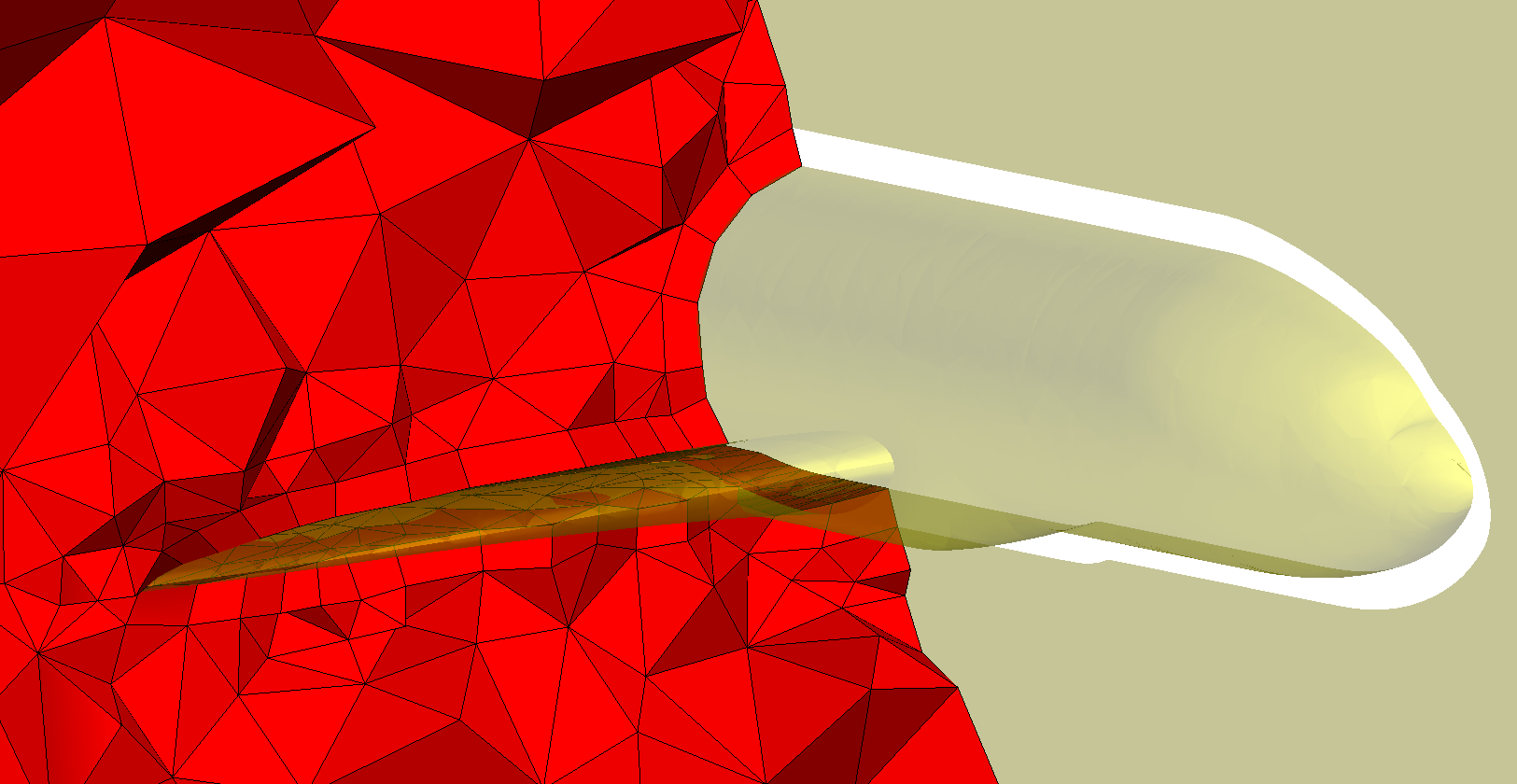} &
        \includegraphics[width=0.49\textwidth]{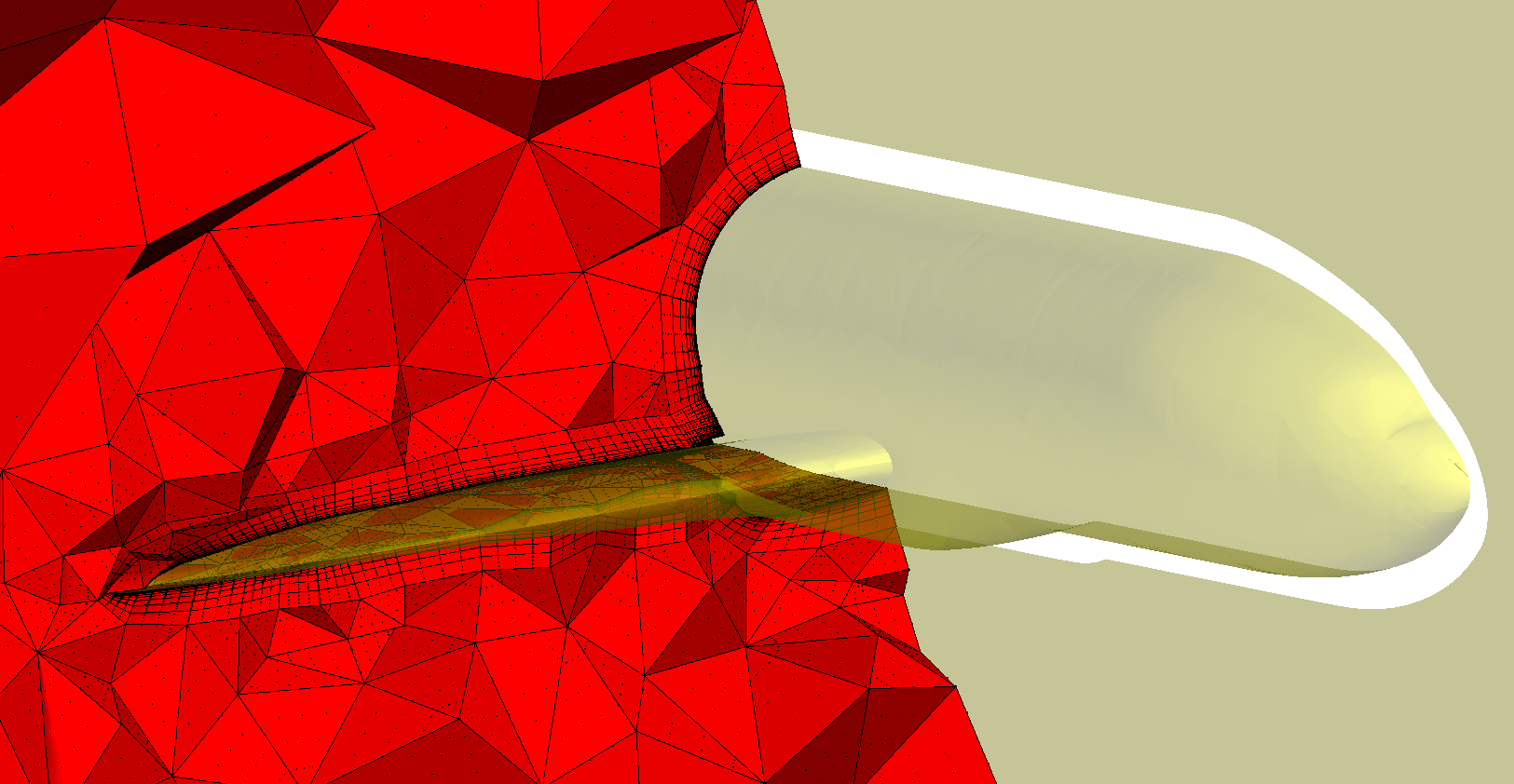} \\
        (a) & (b)
        \end{tabular}
	\caption{A cut view normal to the fuselage of the CRM mesh: (a)
          linear mesh, and (b) high-order mesh.}
	\label{fig:crm_cut}
\end{figure}

\begin{figure}
   \centering
   \begin{tabular}{cc}
     \includegraphics[width=0.45\textwidth]{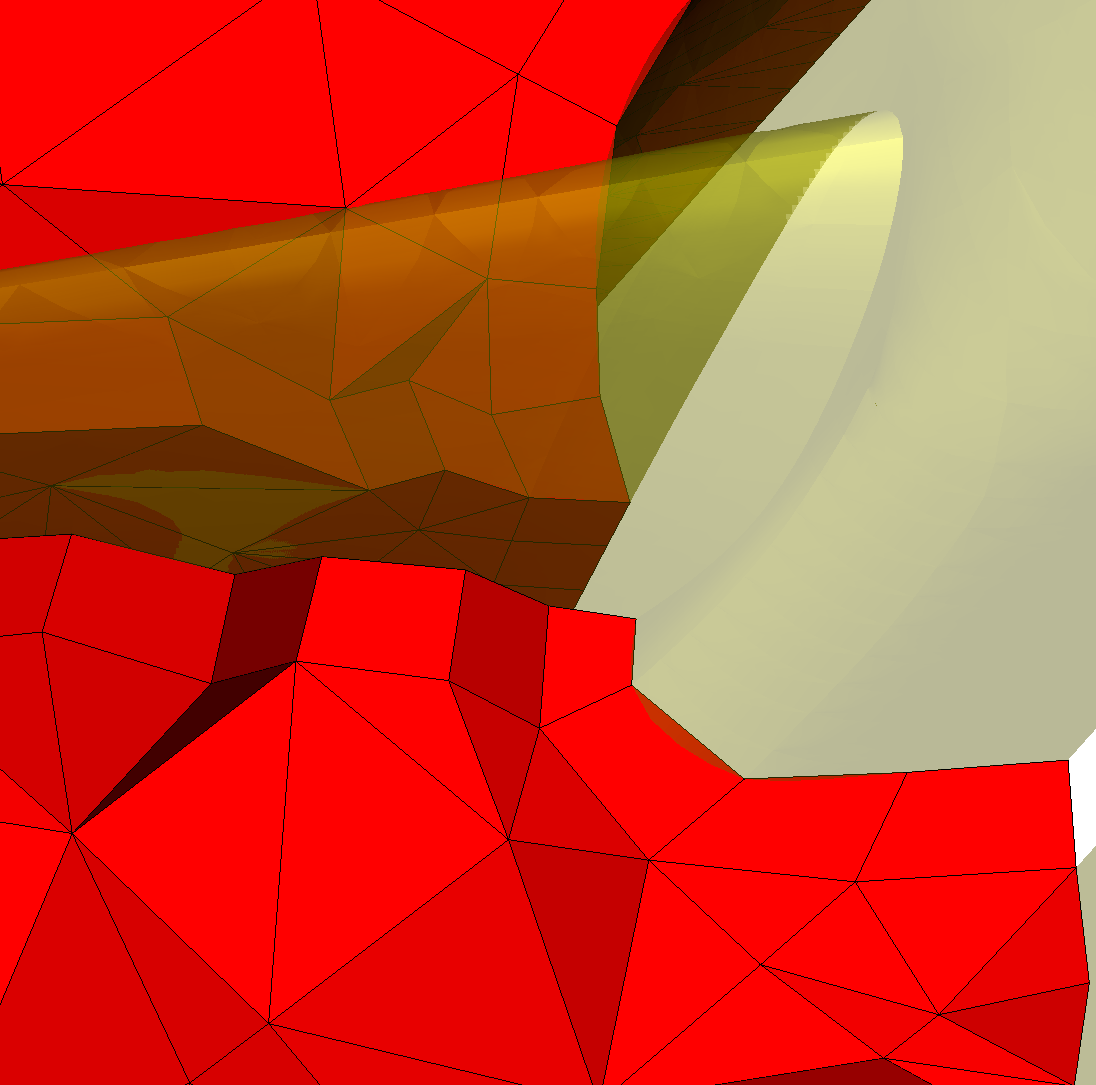} &  
     \includegraphics[width=0.45\textwidth]{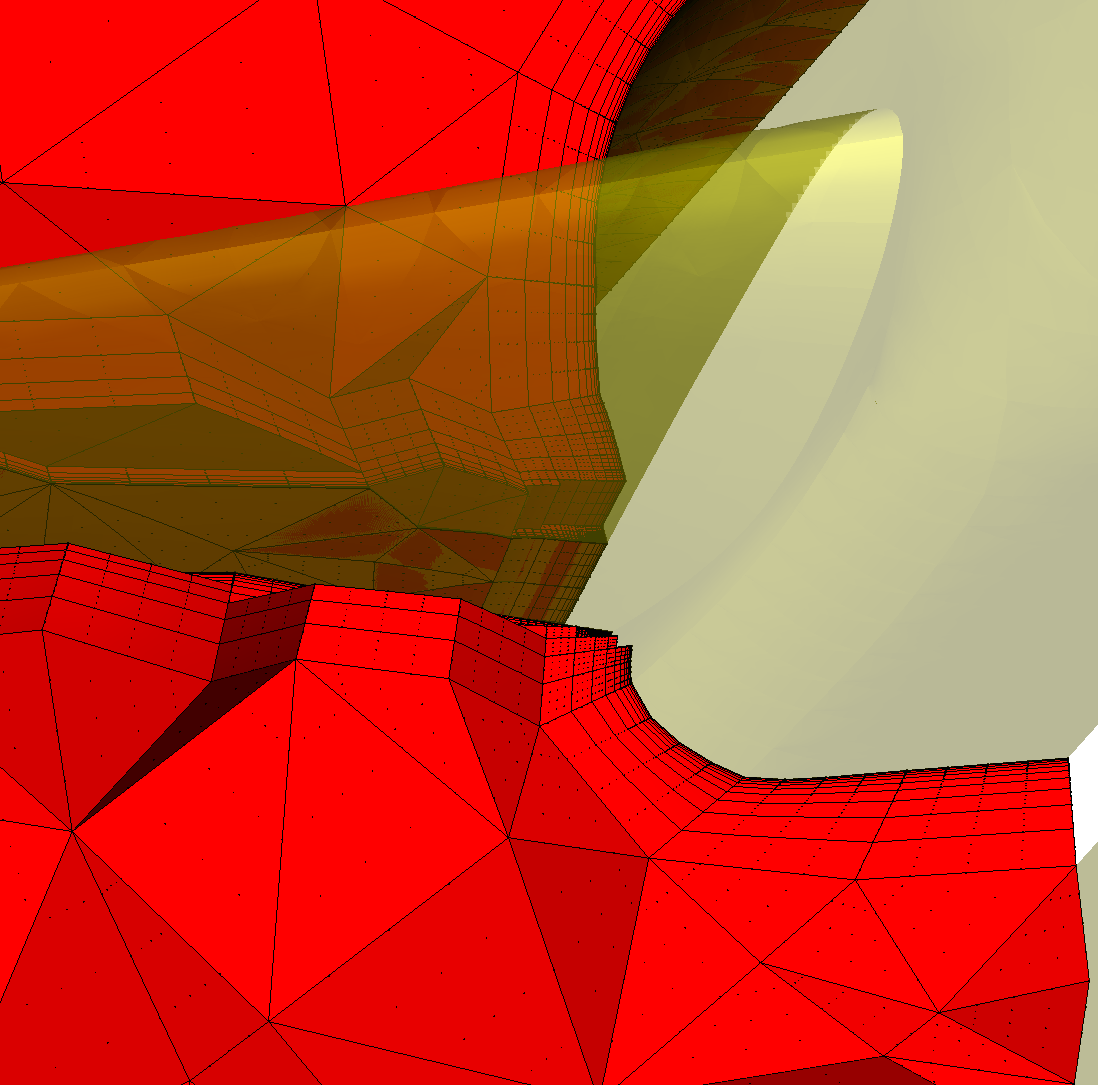} \\
     (a) &  (b)
   \end{tabular}
   \caption{Close-up of the CRM mesh in the region
     adjacent to the wing-fuselage junction: (a) linear mesh, and (b) high-order mesh.}
   \label{fig:crm_bl}
\end{figure}

The medial object decomposition was used to produce an initial coarse linear mesh 
with a single layer of elements in the near-field partition. This boundary-layer
mesh consisted of 33 hexahedra and 2\ 042 prisms. The far-field region was
discretized using 18\ 084 tetrahedra. The characteristics of the linear mesh
can be observed in Fig.~\ref{fig:crm_cut}(a) that shows a cut normal to the 
fuselage through the mesh and in the enlargement of that mesh in the 
wing-fuselage junction of Fig.~\ref{fig:crm_bl}(a).

In the final step of the generation process we apply the isoparametric approach 
to subdivide the coarse boundary-layer mesh into 10 layers with
a progression ratio $r=1.5$. This produces
a high-order mesh with 20\ 420 prisms and 3\ 300 hexahedra with
a maximum element stretching ratio of 60. Views of the cut through the
high-order mesh and the enlargement near the wing-fuselage junction
are shown in Fig.~\ref{fig:crm_cut}(b) and Fig.~\ref{fig:crm_bl}(b), respectively. 

\subsection{NASA Rotor 67}
The geometry considered here is a first stage rotor (NASA Rotor 67) of
a two-stage transonic fan designed and tested at the NASA Glenn
center~\cite{Rotor67-1985}. The original rotor has 22 blades with 
a tip leading-edge radius of 25.7cm and a tip trailing-edge radius 
of 24.25cm. The hub to tip radius ratio is 0.375 at the leading edge
and 0.478 at the trailing edge. Here we consider the geometry for 
a single blade with periodic boundary conditions and without tip 
clearance.  The distinctive feature of this geometry is
the incorporation of periodic features both in the medial object decomposition
and the linear and high-order meshes. 

\begin{figure}
   \centering
   \begin{tabular}{ccc}
     \includegraphics[width=0.3\textwidth]{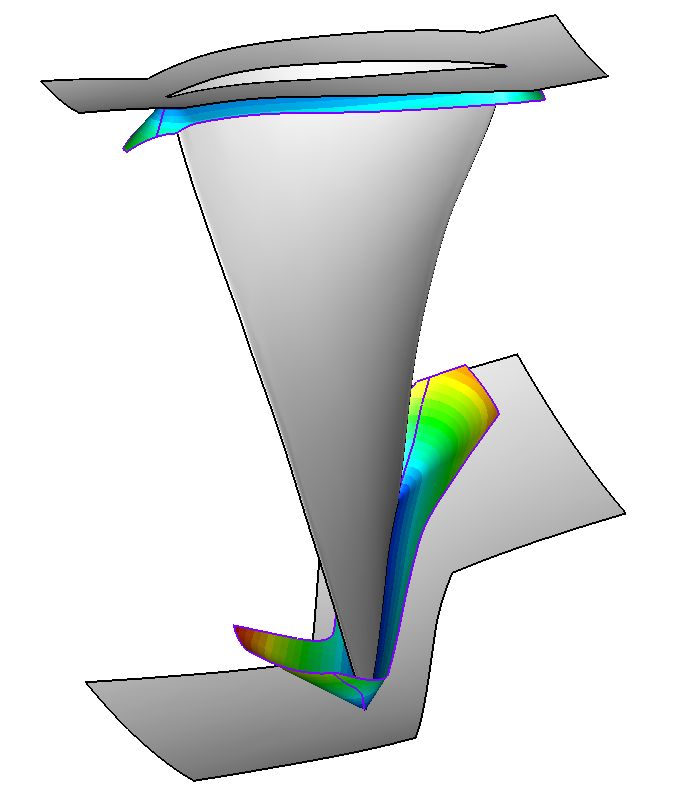} &   
     \includegraphics[width=0.32\textwidth]{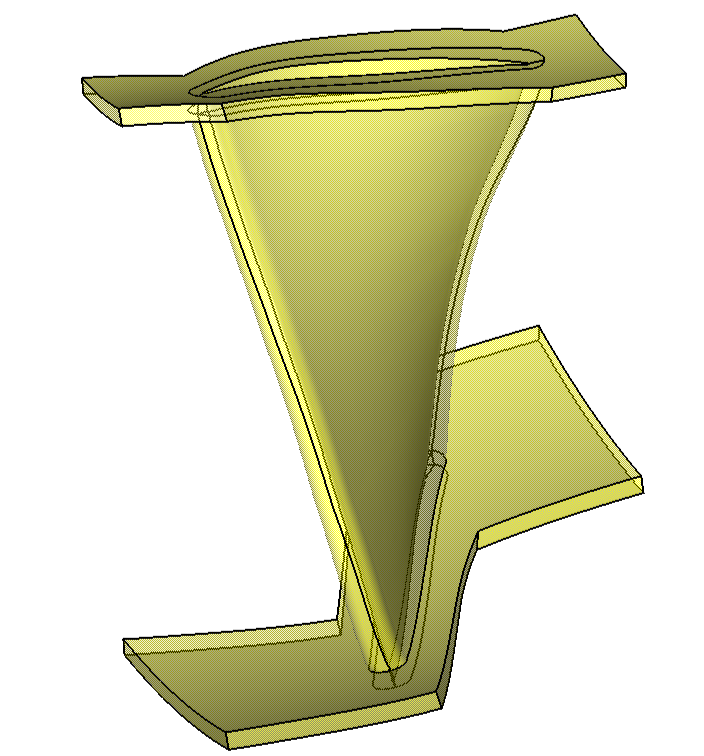} &
     \includegraphics[width=0.26\textwidth]{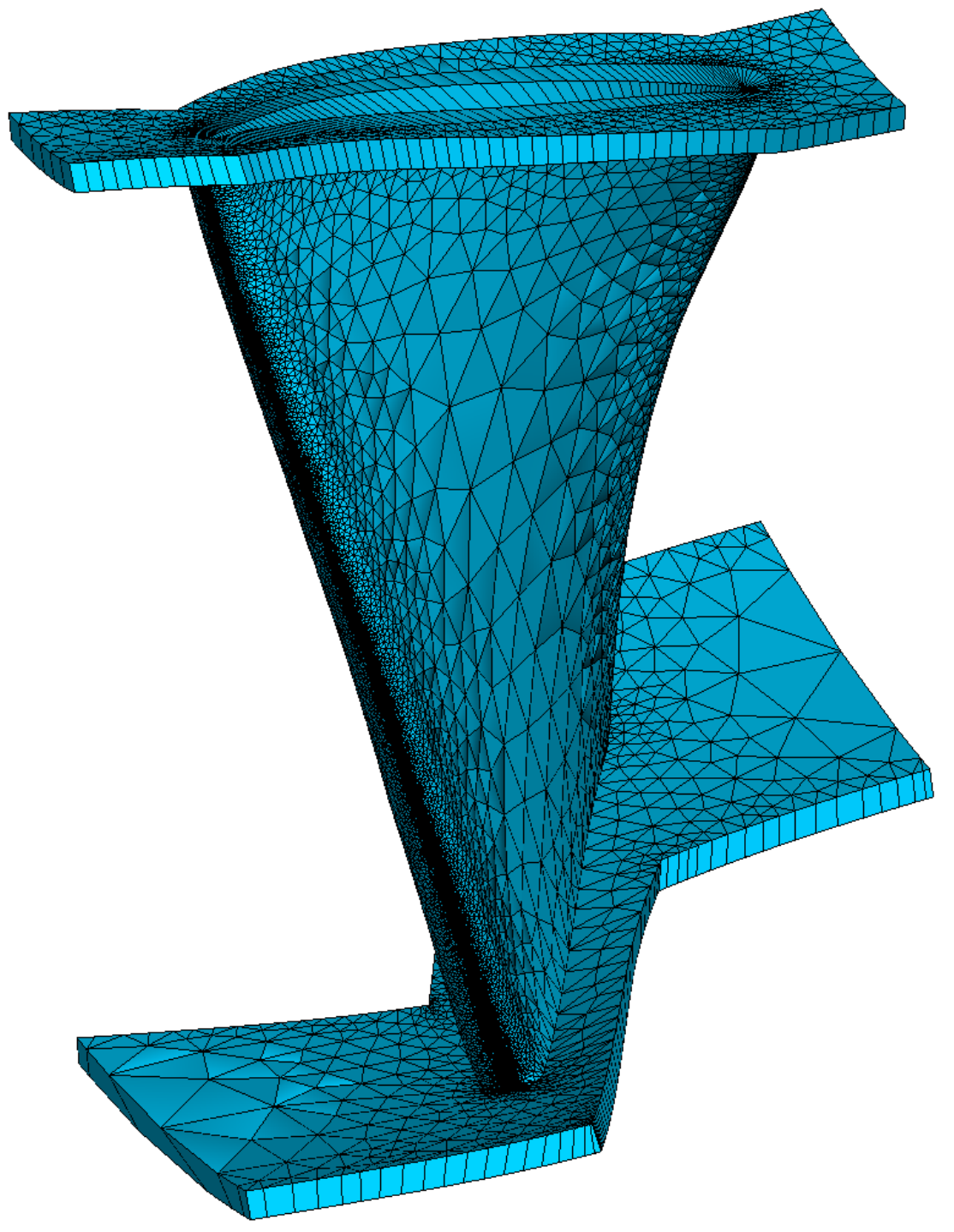}\\
     (a) &  (b) & (c)
   \end{tabular}
   \caption{NASA Rotor 67: (a) Medial object, (b) near-field region and (c) near-field linear mesh.}
   \label{fig:rotor67-2}
\end{figure}

The medial object for this geometry is shown in 
Fig.~\ref{fig:rotor67-2}(a). Fig.~\ref{fig:rotor67-2}(b) depicts the block decomposition
in the near-field region. The linear mesh in that region, which consists of 53\ 830 prisms
and 236 hexahedra, is shown in Fig.~\ref{fig:rotor67-2}(c).

\begin{figure}
   \centering
   \begin{tabular}{cc}
     \includegraphics[width=0.5\textwidth]{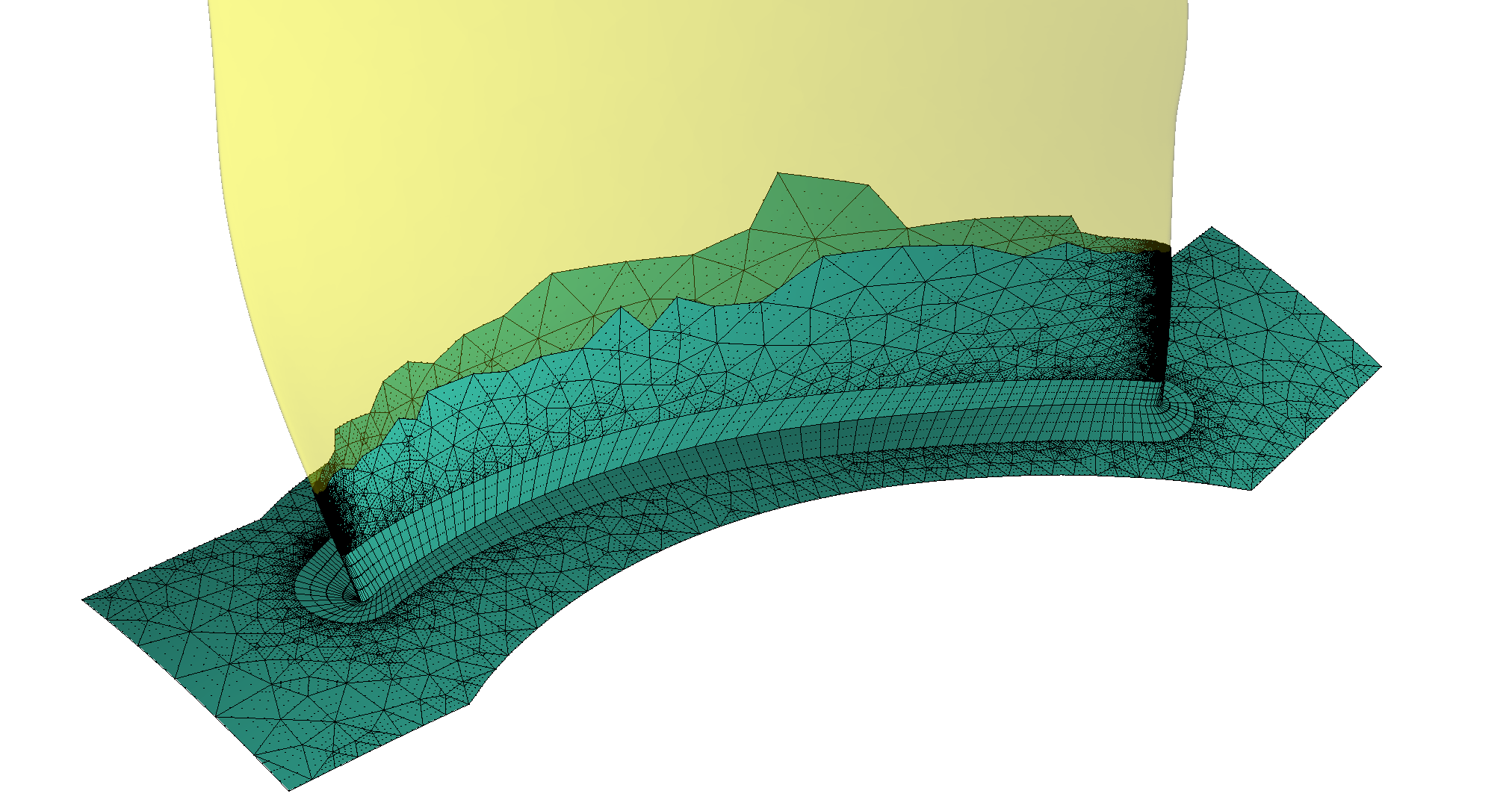} & 
     \includegraphics[width=0.5\textwidth]{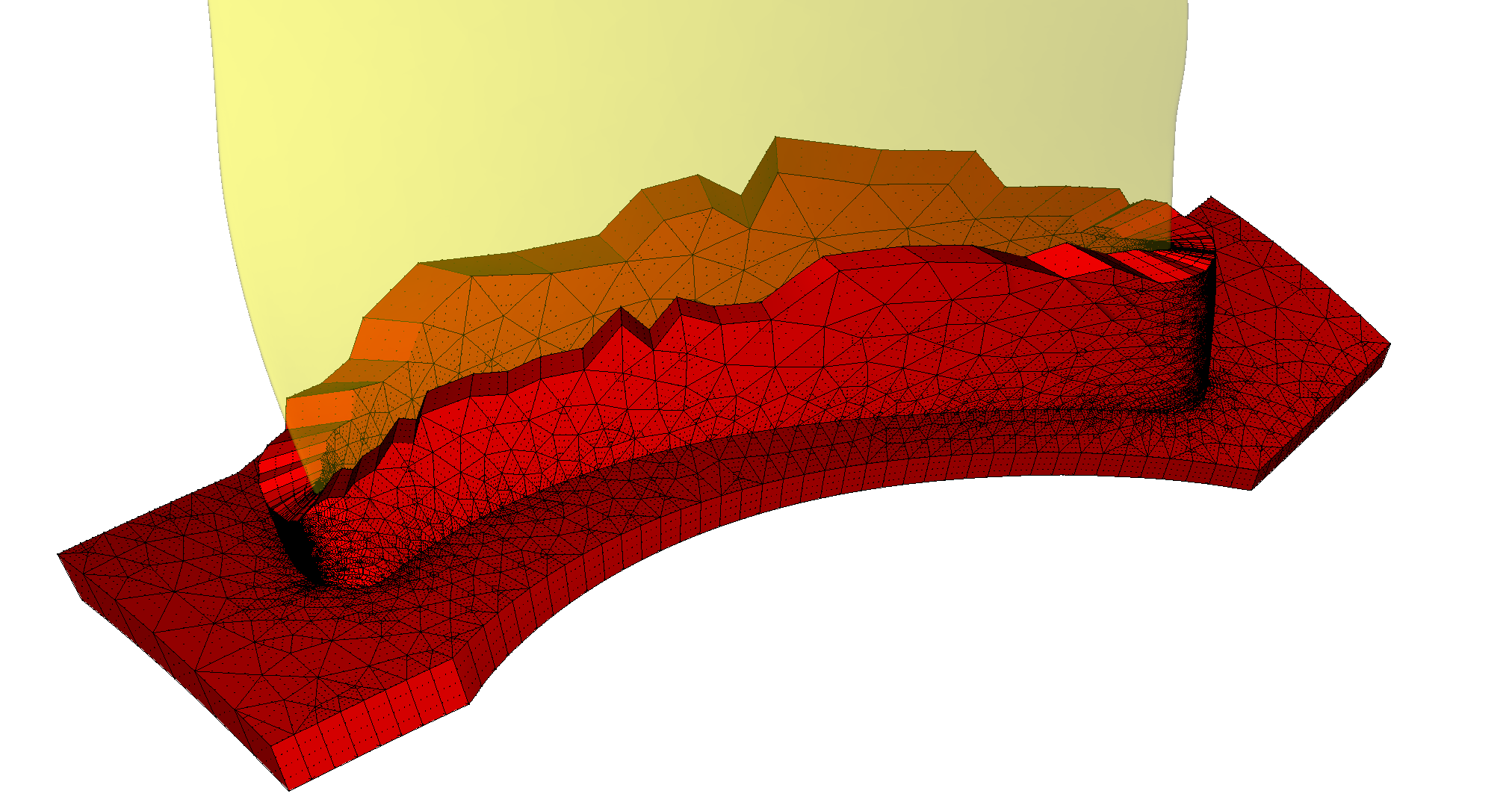} \\
     (a) &  (b)
   \end{tabular}
   \caption{NASA Rotor 67: (a) surface mesh, and (b) boundary-layer mesh.}
   \label{fig:rotor67-3}
\end{figure}

Enlarged views of the surface mesh and the boundary-layer mesh in the vicinity of the blade root are
shown in Fig.~\ref{fig:rotor67-3}(a) and Fig.~\ref{fig:rotor67-3}(b), respectively.  The surface mesh 
contains 53\ 830 triangles and 472 quadrilaterals. 

\begin{figure}
   \centering
     \includegraphics[width=0.5\textwidth]{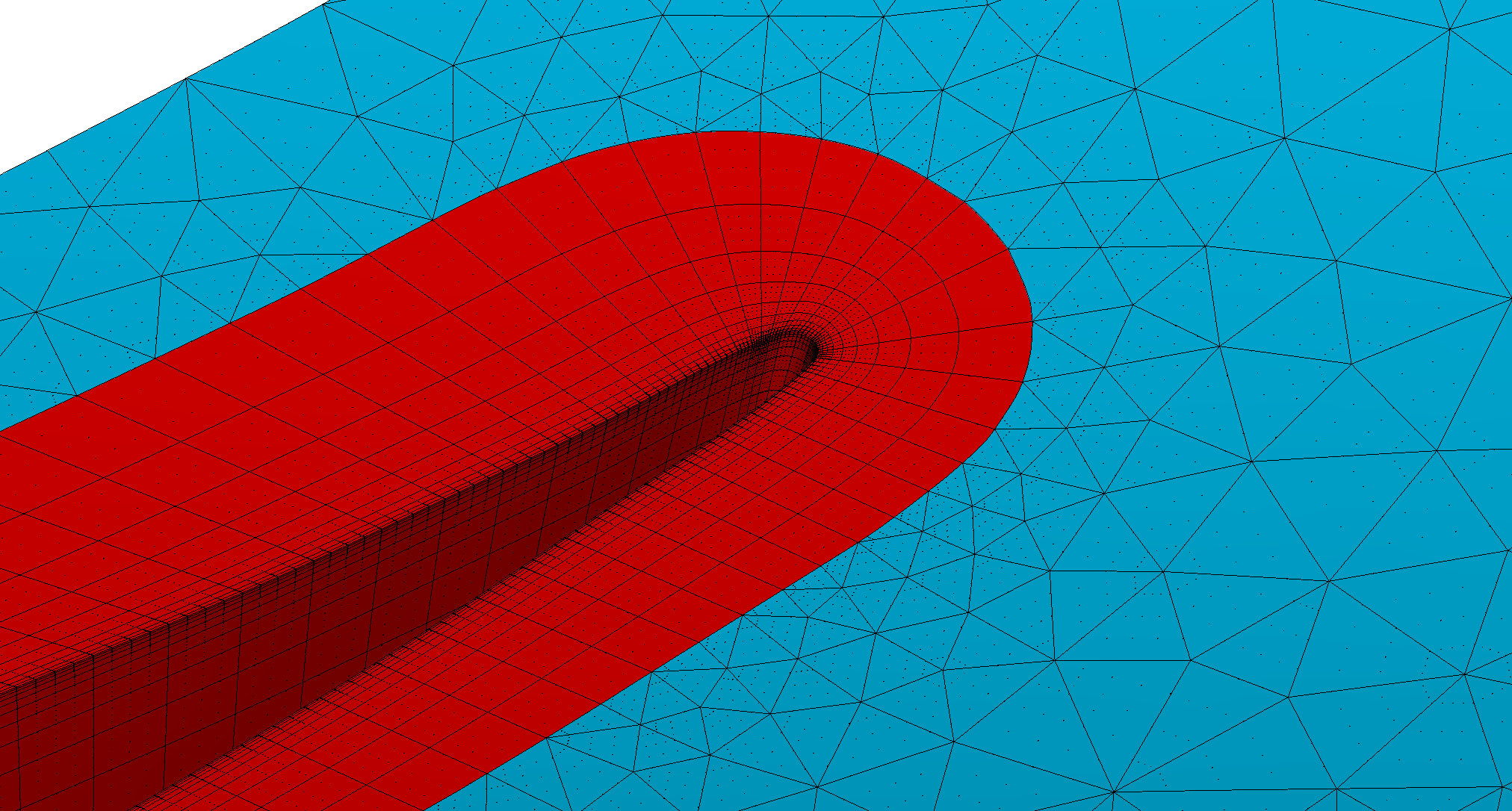}
   \caption{NASA Rotor 67: Enlargement of the root surface mesh near the blade's leading edge.}
   \label{fig:rotor67-4}
\end{figure}

The coarse mixed prismatic-hexahedral linear mesh is split via the isoparametric mapping 
into 10 layers with a progression ratio $r=1.5$ which leads to a maximum stretching ratio of 80 for the
elements near the wall. This linear mesh is then transformed into a high-order mesh
with polynomial order four. A close-up view of the high-order surface mesh near the leading edge of 
the blade's root is shown in Fig.~\ref{fig:rotor67-4}.

\begin{figure}
   \centering
   \begin{tabular}{ccc}
     \includegraphics[width=0.4\textwidth]{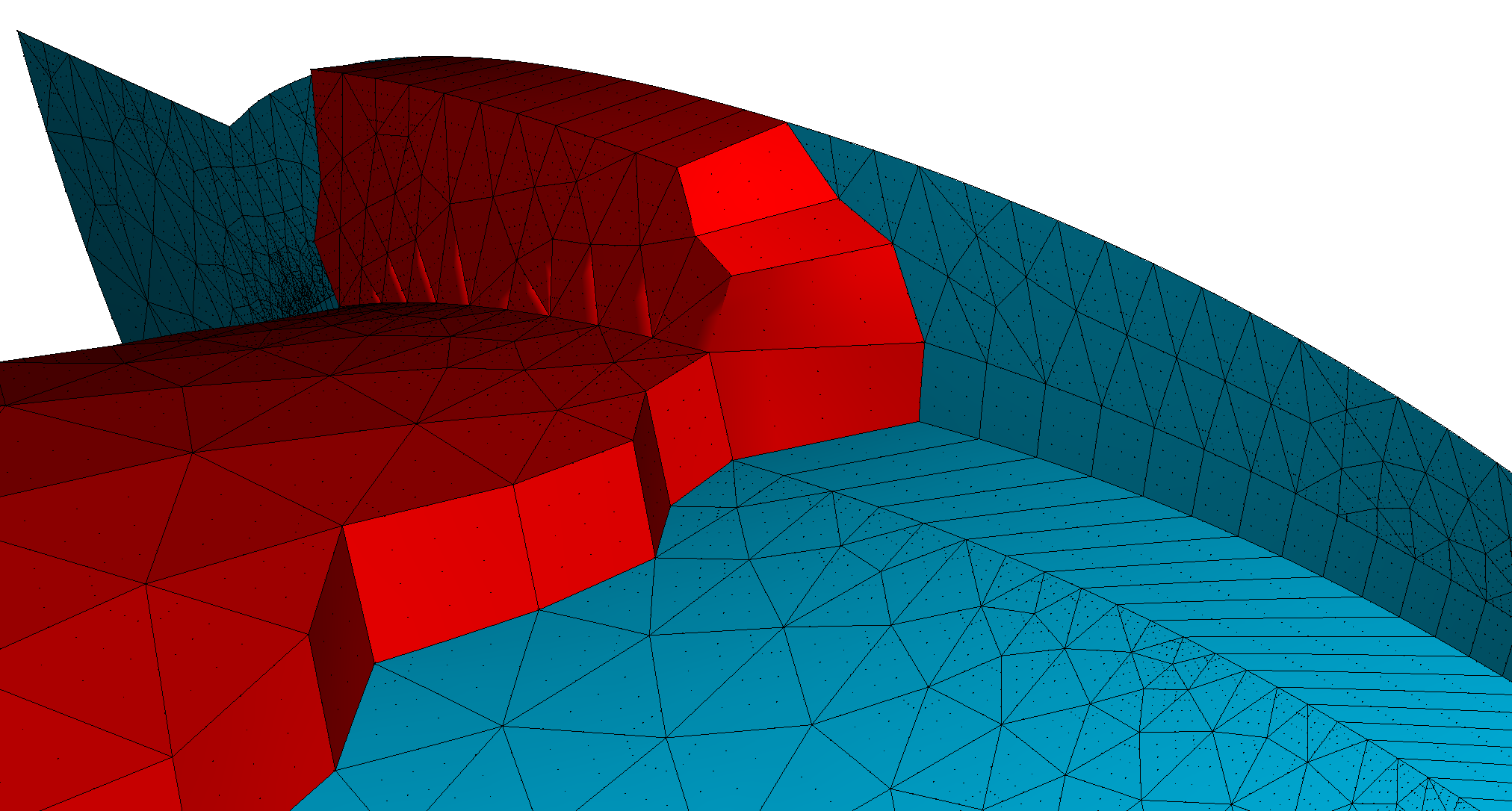} & &  
     \includegraphics[width=0.4\textwidth]{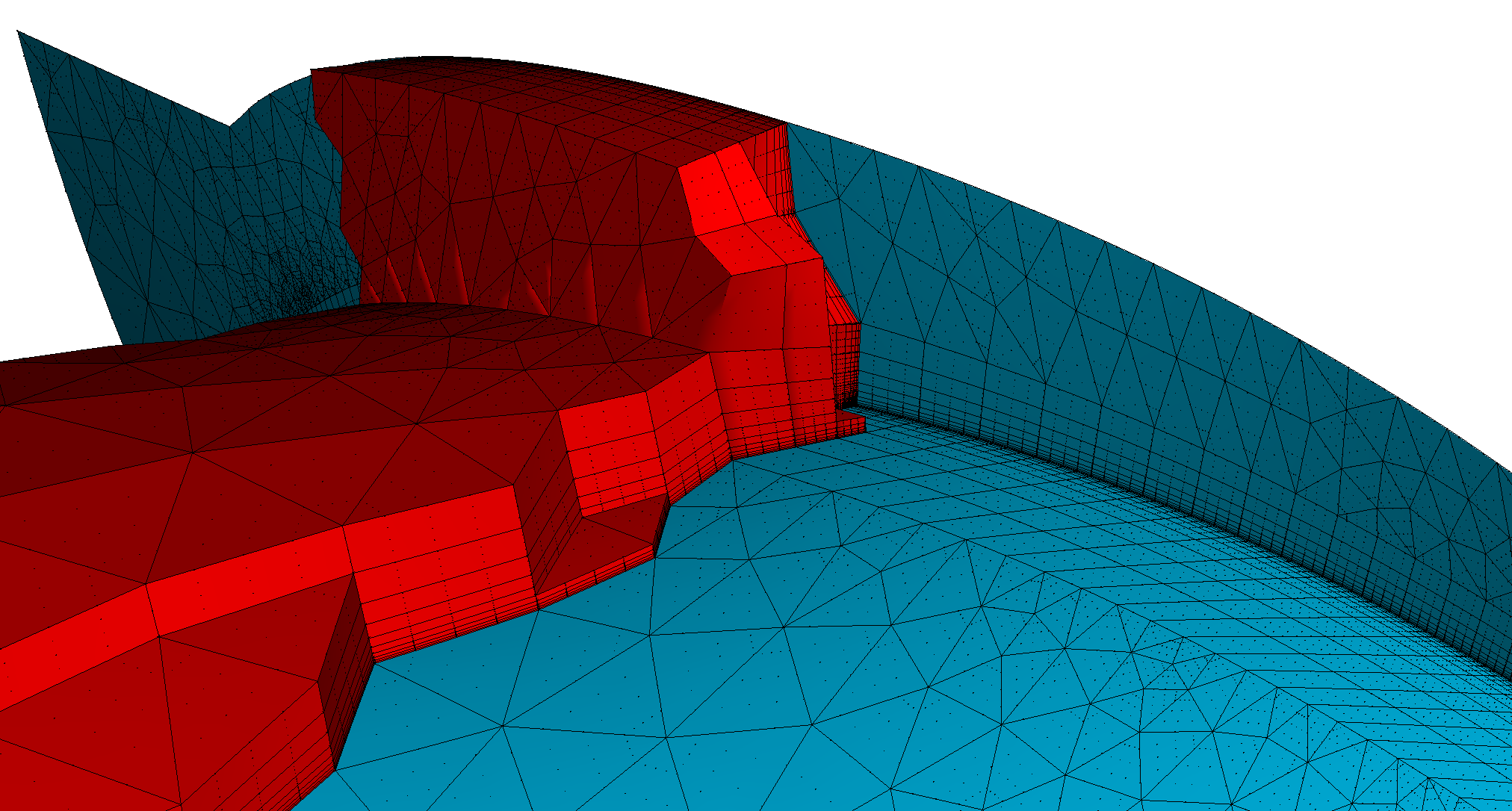} \\
     (a) &  & (b) \\
          \includegraphics[width=0.4\textwidth]{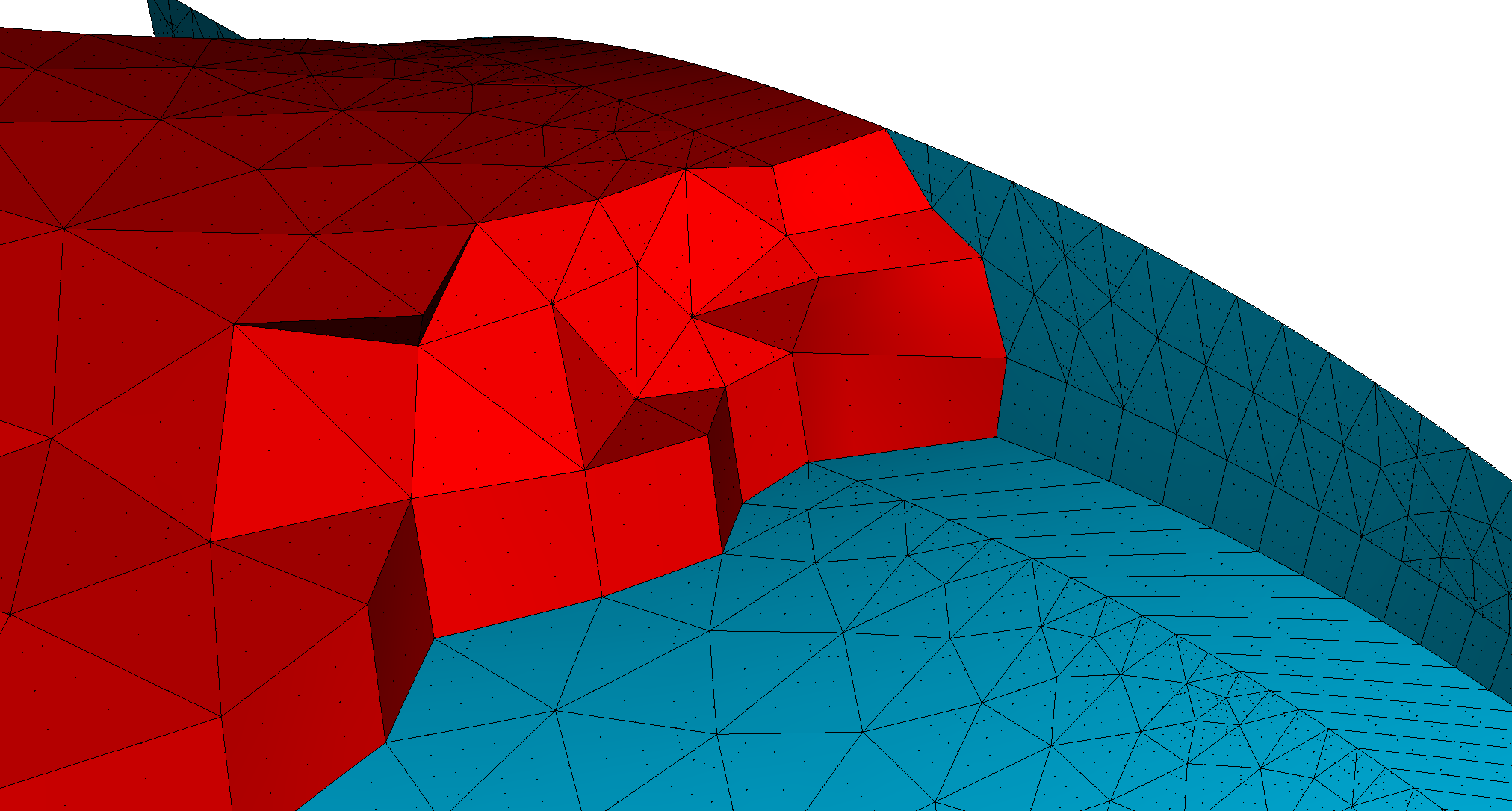} & &  
     \includegraphics[width=0.4\textwidth]{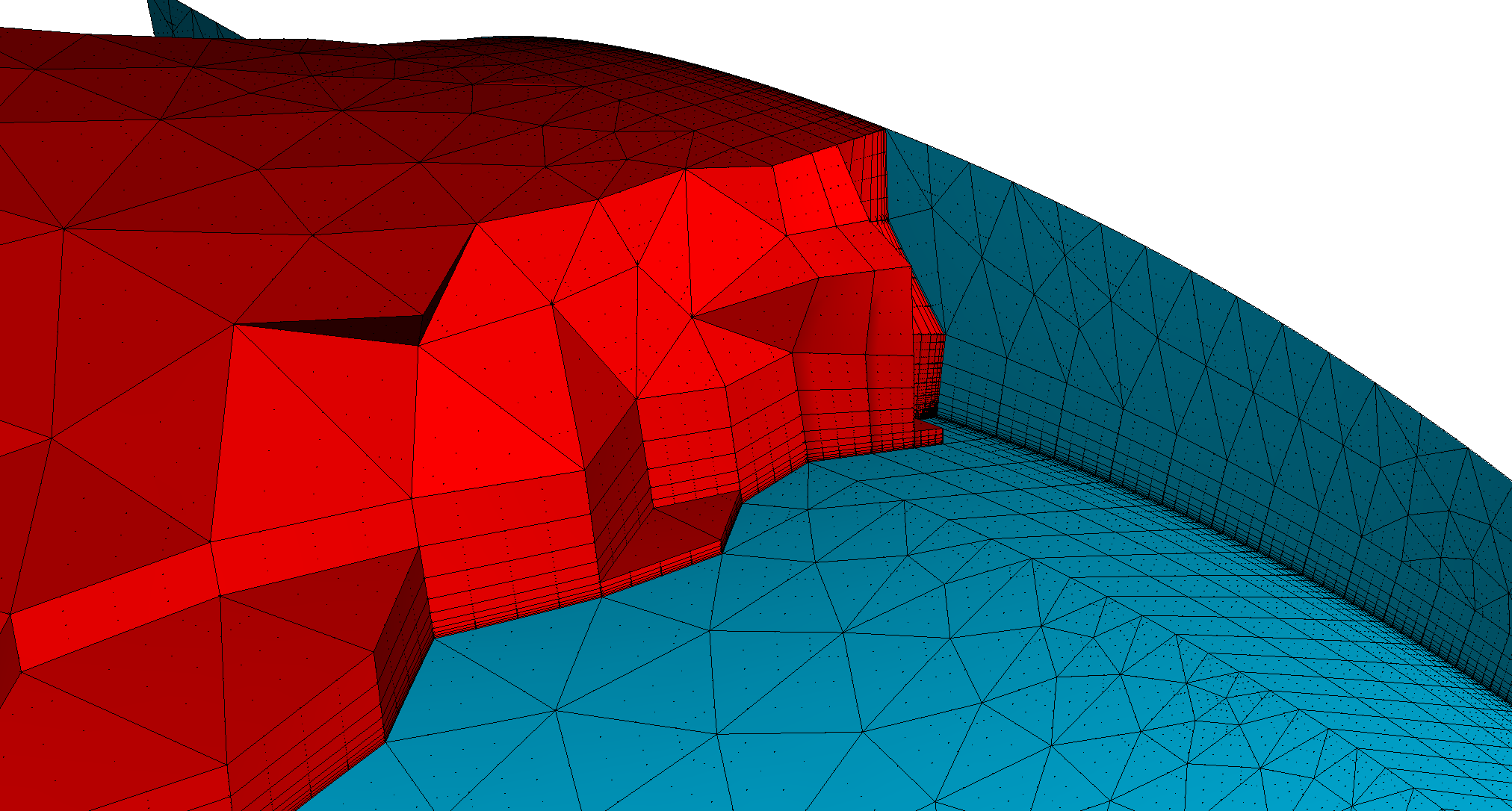} \\
     (c) &  & (d)
   \end{tabular}
   \caption{NASA Rotor 67. Enlargement near the mid-chord of the blade in the near-field region: 
   (a) linear boundary-layer mesh, and (b) high-order boundary-layer mesh. Enlargement of the
   hybrid mesh in the same region: (c) linear mesh and (d) high-order mesh.}
   \label{fig:rotor67-6}
\end{figure}

A pictorial summary of the mesh characteristics of the linear and high-order meshes is given
in Fig.~\ref{fig:rotor67-6} which shows enlargements of these meshes around the mid-chord 
of the blade in the near-field region.

\section{Conclusions}
We have proposed a semi-structured approach for  generating high-order meshes with
high aspect ratio elements to efficiently simulate boundary-layer flows. It 
combines a linear mesh generator for hybrid linear meshes of hexahedral, 
prismatic and tetrahedral elements, based on a medial object approach for
domain decomposition, and an 
\textit{a posteriori} high-order mesh generator.

The domain is decomposed into near-field and far-field regions using
a medial-object approach currently under development within CADfix.
Its parameters have been tuned to produce O-type, described in a 
sister publication \cite{IMR26-2017}, and C-type topologies, described here, 
for the near-field region that facilitate its discretization into meshes of 
prismatic, and of mixed hexahedral and prismatic elements, respectively.
The medial object permits to design near-field regions that allow a
boundary-layer mesh which is  ``thicker" that those achievable by most 
commercially available mesh generators. The far-field region is 
discretized into tetrahedra. Both topologies are simpler to
obtain than a general multi-block decomposition, and yet they are 
sufficiently flexible to deal with reasonably complex geometries. 

There are two main contributions of the paper. The first one is the design of a  
C-type topology for the near-field region in combination with a modified
isoparametric approach that produces hybrid meshes with 
hexahedral elements at junctions. These meshes have a higher quality
and fewer elements than those corresponding to O-type topologies.
The second contribution is the incorporation of periodic surfaces 
both in medial-object decomposition and mesh generation. This permits
the numerical treatment of flow simulations incorporating 
periodic boundary conditions.

The proposed method is robust for both linear and, to some extent, 
high-order meshing, but its ability to produce high-order meshes of 
very good quality depends strongly on the quality of the linear mesh 
and of the distortions induced by the CAD surface mappings. The 
major contributing factors to this problem are the coarseness
of the linear mesh required and the higher sensitivity of
high-order algorithms to distortions in the mappings defining the CAD
curves and surfaces. However, through the use of the semi-structured approach
proposed here and in reference \cite{IMR26-2017} we have improved 
significantly our rate of success at producing high-order meshes 
with complex geometries.

The quality of the mesh can be further enhanced by applying the 
variational approach proposed in reference \cite{turner-2017a} and also
by allowing elemental faces to be curved at the interface between
the near-field and far-field regions. This should be possible since
CADfix represents these as CAD surfaces.


\section*{Acknowledgments}
Mike Turner received financial support from Airbus and EPSRC under an
industrial CASE studentship. This project has also received funding from the
European Union's Horizon 2020 research and innovation programme under
the Marie Sk\l{}odowska-Curie grant agreement No 675008.

\bibliography{Refs}

\begin{thebibliography}{23}
\newcommand{\enquote}[1]{``#1''}
\providecommand{\natexlab}[1]{#1}
\providecommand{\url}[1]{\texttt{#1}}
\providecommand{\urlprefix}{URL }
\expandafter\ifx\csname urlstyle\endcsname\relax
  \providecommand{\doi}[1]{doi:\discretionary{}{}{}#1}\else
  \providecommand{\doi}{doi:\discretionary{}{}{}\begingroup
  \urlstyle{rm}\Url}\fi

\bibitem[{Vincent and Jameson(2011)}]{Vincent+Jameson-2011}
Vincent, P., and Jameson, A., \enquote{Facilitating the Adoption of
  Unstructured High-Order Methods Amongst a Wider Community of Fluid
  Dynamicists,} \emph{Mathematical Modeling Natural of Phenomena}, Vol.~6,
  No.~3, 2011, pp. 97--140.

\bibitem[{Wang et~al.(2013)Wang, Fidkowski, Abgrall, Bassi, Caraeni, Cary,
  Deconinck, Hartmann, Hillewaert, Huynh, Kroll, May, Persson, van Leer, and
  Visbal}]{Wang-2013}
Wang, Z.~J., Fidkowski, K., Abgrall, R., Bassi, F., Caraeni, D., Cary, A.,
  Deconinck, H., Hartmann, R., Hillewaert, K., Huynh, H.~T., Kroll, N., May,
  G., Persson, P.-O., van Leer, B., and Visbal, M., \enquote{High-order {CFD}
  methods: {C}urrent status and perspective,} \emph{International Journal for
  Numerical Methods in Fluids}, Vol.~72, No.~8, 2013, pp. 811--845.

\bibitem[{Armstrong et~al.(2015)Armstrong, Fogg, Tierney, and
  Robinson}]{Armstrong-2015}
Armstrong, C.~G., Fogg, H.~J., Tierney, C.~M., and Robinson, T.~T.,
  \enquote{Common Themes in Multi-block Structured Quad/Hex Mesh Generation,}
  \emph{Procedia Engineering}, Vol. 124, 2015, pp. 70--82.

\bibitem[{Turner et~al.(2017{\natexlab{a}})Turner, Peir\'o, and
  Moxey}]{turner-2017a}
Turner, M., Peir\'o, J., and Moxey, D., \enquote{Curvilinear mesh generation
  using a variational framework,} \emph{Computer Aided Design},
  2017{\natexlab{a}}.
\newblock \doi{doi.org/10.1016/j.cad.2017.10.004}, available online.

\bibitem[{ITI-Global(2017)}]{CADfix}
ITI-Global, \enquote{{CADfix}: {CAD} translation, healing, repair, and
  transformation,} , 2017.

\bibitem[{Cantwell et~al.(2015)Cantwell, Moxey, Comerford, Bolis, Rocco,
  Mengaldo, de~Grazia, Yakovlev, Lombard, Ekelschot, Jordi, Xu, Mohamied,
  Eskilsson, Nelson, Vos, Biotto, Kirby, and Sherwin}]{cantwell-2015}
Cantwell, C.~D., Moxey, D., Comerford, A., Bolis, A., Rocco, G., Mengaldo, G.,
  de~Grazia, D., Yakovlev, S., Lombard, J.-E., Ekelschot, D., Jordi, B., Xu,
  H., Mohamied, Y., Eskilsson, C., Nelson, B., Vos, P., Biotto, C., Kirby,
  R.~M., and Sherwin, S.~J., \enquote{Nektar++: An open-source spectral/hp
  element framework,} \emph{Computer Physics Communications}, Vol. 192, 2015,
  pp. 205--219.
\newblock \doi{10.1016/j.cpc.2015.02.008}.

\bibitem[{Sherwin and Peir\'{o}(2002)}]{Sherwin+Peiro-2002}
Sherwin, S., and Peir\'{o}, J., \enquote{Mesh generation in curvilinear domains
  using high-order elements,} \emph{International Journal for Numerical Methods
  in Engineering}, Vol.~53, 2002, pp. 207--223.

\bibitem[{Moxey et~al.(2015{\natexlab{a}})Moxey, Green, Sherwin, and
  Peir{\'o}}]{moxey-2015a}
Moxey, D., Green, M.~D., Sherwin, S.~J., and Peir{\'o}, J., \enquote{An
  isoparametric approach to high-order curvilinear boundary-layer meshing,}
  \emph{Comput. Meth. Appl. Mech. Eng.}, Vol. 283, 2015{\natexlab{a}}, pp.
  636--650.

\bibitem[{Ope(Accessed December 2017)}]{OpenCascade}
\enquote{{OPEN CASCADE},} https://www.opencascade.com, Accessed December 2017.

\bibitem[{Bucklow et~al.(2017)Bucklow, Fairey, and Gammon}]{Bucklow-2017}
Bucklow, J.~H., Fairey, R., and Gammon, M.~R., \enquote{An automated workflow
  for high quality {CFD} meshing using the {3D} medial object,} \emph{23rd AIAA
  Computational Fluid Dynamics Conference}, Denver, Colorado, 2017.
\newblock AIAA 2017-3454.

\bibitem[{Chow et~al.(1997)Chow, Zilliac, and Bradshaw}]{Chow-1997}
Chow, S., Zilliac, G., and Bradshaw, P., \enquote{Mean and turbulence
  measurements in the near field of a wingtip vortex,} \emph{AIAA Journal},
  Vol.~35, No.~10, 1997, pp. 1561--1567.

\bibitem[{Lombard et~al.(2016)Lombard, Moxey, Sherwin, Hoessler, Dhandapani,
  and Taylor}]{Lombard-2016}
Lombard, J.-E.~W., Moxey, D., Sherwin, S.~J., Hoessler, J. F.~A., Dhandapani,
  S., and Taylor, M.~J., \enquote{Implicit large-eddy simulation of a wingtip
  vortex,} \emph{AIAA Journal}, Vol.~54, No.~2, 2016, pp. 506--518.

\bibitem[{Blum(1967)}]{Blum-1967}
Blum, H., \enquote{A transformation for extracting new descriptors of shape,}
  \emph{Models for the Perception of Speech and Visual Form}, Vol.~5, 1967, pp.
  362--380.

\bibitem[{Sheehy(1994)}]{Sheehy-1994}
Sheehy, D., \enquote{Medial surface computation using a domain {D}elaunay
  triangulation,} Ph.D. thesis, Queen's University of Belfast, 1994.

\bibitem[{Bucklow(2014)}]{Bucklow-2014}
Bucklow, H., \enquote{{3D} medial object computation using a domain {D}elaunay
  triangulation,} , 2014.
\newblock In Medial Object Technology Workshop.

\bibitem[{Tam and Armstrong(1993)}]{Tam-1993}
Tam, T., and Armstrong, C.~G., \enquote{Finite element mesh control by integer
  programming,} \emph{International Journal for Numerical Methods in
  Engineering}, Vol.~36, No.~15, 1993, pp. 2581--2605.

\bibitem[{lps(2017)}]{lpsolve-2017}
\enquote{Software Package \texttt{lpsolve}, Ver. 5.5,}
  http://lpsolve.sourceforge.net, 2017.

\bibitem[{Byrd et~al.(1995)Byrd, Lu, Nocedal, and Zhu}]{Byrd-1995}
Byrd, R.~H., Lu, P., Nocedal, J., and Zhu, C., \enquote{A limited memory
  algorithm for bound constrained optimisation,} \emph{SIAM Journal on
  Scientific Computing}, Vol.~16, No.~5, 1995, pp. 1190--1208.

\bibitem[{Moxey et~al.(2015{\natexlab{b}})Moxey, Green, Sherwin, and
  Peir\'o}]{moxey-2015d}
Moxey, D., Green, M.~D., Sherwin, S.~J., and Peir\'o, J., \emph{On the
  generation of curvilinear meshes through subdivision of isoparametric
  elements}, Springer, 2015{\natexlab{b}}, pp. 203--215.

\bibitem[{Vassberg et~al.(2008)Vassberg, DeHaan, Rivers, and Wahls}]{CRM-2008}
Vassberg, J.~C., DeHaan, M.~A., Rivers, S.~M., and Wahls, R.~A.,
  \enquote{Development of a Common Research Model for Applied {CFD} Validation
  Studies,} \emph{26th AIAA Applied Aerodynamics Conference}, Honolulu, Hawaii,
  2008.
\newblock AIAA-2008-6919.

\bibitem[{CRM(Accessed December 2017)}]{CRM-step}
\enquote{{Common Research Model geometry repository},}
  https://commonresearchmodel.larc.nasa.gov, Accessed December 2017.

\bibitem[{Pierzga and Wood(1985)}]{Rotor67-1985}
Pierzga, M., and Wood, J., \enquote{Investigation of the Three-Dimensional Flow
  Field Within a Transonic Fan Rotor: {E}xperiment and Analysis,} \emph{ASME
  Journal of Engineering for Gas Turbines and Power}, Vol. 107, No.~2, 1985,
  pp. 436--449.

\bibitem[{Turner et~al.(2017{\natexlab{b}})Turner, Moxey, Peir{\'o}, Gammon,
  Pollard, and Bucklow}]{IMR26-2017}
Turner, M., Moxey, D., Peir{\'o}, J., Gammon, M., Pollard, C.~R., and Bucklow,
  H., \enquote{A framework for the generation of high-order curvilinear hybrid
  meshes for {CFD} simulations,} \emph{Procedia Engineering}, Vol. 203,
  2017{\natexlab{b}}, pp. 206--218.

\end{thebibliography}

\end{document}